\documentclass[final,3p,times,authoryear]{elsarticle} 

\usepackage{amssymb}
\usepackage{amsmath}

\usepackage[colorlinks]{hyperref}

\usepackage{lscape}

\usepackage[usenames]{color}
\usepackage[normalem]{ulem}

\newcommand{\srim}{{\tt SRIM}}

\newcommand{\talys}{{\tt TALYS}}

\newcommand{\mcnp}{{\tt MCNP6}}
\newcommand{\mycode}{{\tt argon39}}
\newcommand{\drm}{\mathrm{d}}
\newcommand{\td}[2]{\frac{\drm#1}{\drm#2}}
\newcommand{\tdl}[2]{\drm#1/\drm#2}
\newcommand{\isot}[2]{\ensuremath{{^{#2}\text{#1}}}}

\newcommand{\HEfor}{\isot{He}{4}}
\newcommand{\Oeit}{\isot{O}{18}}
\newcommand{\NEone}{\isot{Ne}{21}}
\newcommand{\MGfor}{\isot{Mg}{24}}
\newcommand{\Knine}{\isot{K}{39}}
\newcommand{\Kfort}{\isot{K}{40}}
\newcommand{\ARnine}{\isot{Ar}{39}}
\newcommand{\ARfort}{\isot{Ar}{40}}
\newcommand{\ARfortS}{\ensuremath{\isot{Ar}{40}^*}}
\newcommand{\THtwo}{\isot{Th}{232}}
\newcommand{\Ufive}{\isot{U}{235}}
\newcommand{\Ueit}{\isot{U}{238}}
\newcommand{\cfac}{\ensuremath{\mathcal{C}}}
\newcommand{\nalf}{\ensuremath{n_\alpha}}
\newcommand{\Salf}{\ensuremath{S_\alpha}}
\newcommand{\Snine}{\ensuremath{S_{39\text{Ar}}}}
\newcommand{\Sone}{\ensuremath{S_{21\text{Ne}}}}
\newcommand{\Sneut}{\ensuremath{S_n}}
\newcommand{\ThU}{\ensuremath{\frac{A_\text{Th}}{A_\text{U}}}}
\newcommand{\Yan}{\ensuremath{Y_{\alpha,n}}}
\newcommand{\Ysf}{\ensuremath{Y_\text{SF}}}
\newcommand{\Ea}{\ensuremath{E_\alpha}}
\newcommand{\Eaz}{\ensuremath{E_{\alpha_0}}}
\newcommand{\Eth}{\ensuremath{E_{th}}}

\newcommand{\Vcoul}{\ensuremath{V_C}}
\newcommand{\an}{\ensuremath{(\alpha,n)}}
\newcommand{\np}{\ensuremath{(n,p)}}
\newcommand{\na}{\ensuremath{(n,\alpha)}}
\newcommand{\CSan}{\ensuremath{\sigma_{\alpha,n}}}
\newcommand{\ncms}{n\,cm$^{-2}$\,s$^{-1}$}

\newcommand{\Pnine}{\ensuremath{^{39}P}}

\newcommand{\A}{\ensuremath{\alpha}}
\newcommand{\phiR}{\ensuremath{\phi_\text{R}}}

\newcommand{\RR}{\ensuremath{R_\text{R}}}
\newcommand{\RF}{\ensuremath{R_\text{F}}}
\newcommand{\Ra}{\ensuremath{R_a}}

\journal{Geochimica et Cosmochimica Acta}

\begin{document}

\begin{frontmatter}



\title{Subterranean production of neutrons, {\ARnine} and {\NEone}: Rates and uncertainties}


\author[adr1]{Ond\v rej \v Sr\'amek\corref{cor1}}
\ead{ondrej.sramek@gmail.com}
\cortext[cor1]{Principal corresponding author.}
\author[adr2]{Lauren Stevens}
\author[adr2,adr3]{William F. McDonough\corref{cor2}}
\ead{mcdonoug@umd.edu}
\cortext[cor2]{Corresponding author.}
\author[adr4]{Sujoy Mukhopadhyay}
\author[adr5]{R. J. Peterson}

\address[adr1]{Department of Geophysics, Faculty of Mathematics and Physics, Charles University, V Hole\v sovi\v ck\'ach 2, 18000 Praha 8, Czech Republic}
\address[adr2]{Department of Chemistry and Biochemistry, University of Maryland, College Park, MD 20742, United States}
\address[adr3]{Department of Geology, University of Maryland, College Park, MD 20742, United States}
\address[adr4]{Department of Earth and Planetary Sciences, University of California Davis, Davis, CA 95616, United States}
\address[adr5]{Department of Physics, University of Colorado Boulder, Boulder, CO 80309-0390, United States}

\begin{abstract}
Accurate understanding of the subsurface production rate of the radionuclide {\ARnine} is necessary for argon dating techniques and noble gas geochemistry of the shallow and the deep Earth, and is also of interest to the WIMP dark matter experimental particle physics community. Our new calculations of subsurface production of neutrons, {\NEone}, and {\ARnine} take advantage of the state-of-the-art reliable tools of nuclear physics to obtain reaction cross sections and spectra (\talys) and to evaluate neutron propagation in rock (\mcnp). We discuss our method and results in relation to previous studies and show the relative importance of various neutron, {\NEone}, and {\ARnine} nucleogenic production channels. Uncertainty in nuclear reaction cross sections, which is the major contributor to overall calculation uncertainty, is estimated from variability in existing experimental and library data. Depending on selected rock composition, on the order of $10^7$--$10^{10}$ $\alpha$ particles are produced in one kilogram of rock per year (order of 1--$10^3$\,kg$^{-1}$\,s$^{-1}$); the number of produced neutrons is lower by $\sim6$ orders of magnitude, {\NEone} production rate drops by an additional factor of 15--20, and another one order of magnitude or more is dropped in production of {\ARnine}. Our calculation yields a nucleogenic {\NEone/\HEfor} production ratio of $(4.6\pm0.6)\times10^{-8}$ in Continental Crust and $(4.2\pm0.5)\times10^{-8}$ in Oceanic Crust and Depleted Mantle. Calculated {\ARnine} production rates span a great range from $29\pm9$\,atoms\,kg-rock$^{-1}$\,yr$^{-1}$ in the K--Th--U-enriched Upper Continental Crust to $(2.6\pm0.8)\times10^{-4}$\,atoms\,kg-rock$^{-1}$\,yr$^{-1}$ in Depleted Upper Mantle. 
Nucleogenic {\ARnine} production exceeds the cosmogenic production below $\sim700$ meters depth and thus, affects radiometric ages of groundwater. The {\ARnine} chronometer, which fills in a gap between \isot{H}{3} and \isot{C}{14}, is particularly important given the need to tap deep reservoirs of ancient drinking water.
\end{abstract}


\end{frontmatter}

\tableofcontents


\section{Introduction}

Argon-39 is a noble gas radionuclide with half-life of $269\pm3$\,years \citep{NNDC} used in dating of hydrological and geological processes with timescales of a few tens to about 1000\,years \citep{loosli:1983,ballentine:2002,lu:2014,yokochi:2012,yokochi:2013}. Accurate estimates of production rates of noble gas isotopes, such as {\ARnine}, {\ARfort} and {\NEone}, in rocks are necessary for dating groundwaters and in proper interpretation of isotopic signatures of gases originating in Earth's interior \citep[e.g.,][]{graham:2002,tucker:2014}.

Production of {\ARnine} in the Earth's atmosphere is dominated by the cosmogenic production channel, where {\ARnine} is produced from the abundant stable {\ARfort} ($\sim1$\,wt.\% of atmosphere) by the cosmic ray-induced reaction $\ARfort(n,2n)\ARnine$ \citep[e.g.,][]{loosli:1968}. The notation is the standard shorthand for a transfer reaction $\ARfort + n \rightarrow 2n + \ARnine$ and $n$ stands for a neutron. The {\ARnine} cosmogenic production keeps the atmospheric {\ARnine/Ar} mass ratio at a present-day steady-state value of $(8.0\pm0.6)\times10^{-16}$ \citep{benetti:2007}.

Below the Earth's surface {\ARnine} can also be produced cosmogenically, by cosmic ray negative muon interactions, in particular by negative muon capture on {\Knine} (93.3\,\% of all potassium; $\mu^- + \Knine \rightarrow \nu_\mu + \ARnine$). As the secondary cosmic ray muon flux drops off with depth, at depths greater than 2000\,meters water equivalent (m.w.e.; $\sim700$\,m in the rock) the nucleogenic production by {\Knine\np\ARnine} dominates \citep{mei:2010}.

Argon extracted from several deep natural gas reservoirs shows \ARnine/Ar ratios below the atmospheric value \citep{acostakane:2008,xucalaprice:2012arxiv}. For this reason, underground gas sources have attracted the attention of astroparticle physics projects that require a source of Ar with minimal radioactive {\ARnine} for low-energy rare event detection of WIMP Dark Matter interactions and neutrinoless double beta decay using liquid argon detectors.

This paper provides an evaluation of the {\ARnine} nucleogenic production rate for specified rock compositions by calculating the rate of  {\ARnine} atoms produced by naturally occurring fast neutrons. The calculation also yields results for other nuclides, in particular the rare stable {\NEone} (0.27\,\% of neon). 
We describe in detail the methods of calculation of neutron, {\NEone} and {\ARnine} yields and production rates, including the use of nuclear physics software tools (\srim, \talys, \mcnp), and the input data, with the goal to present reproducible results. In section \ref{sec:overview} we give the overview of the calculation. Section \ref{sec:alpha} deals with $\alpha$-particle production. Section \ref{sec:neutron} discusses production of neutrons, both by {\an} reactions and by spontaneous fission. Neutron-induced reactions, in particular {\Knine\np\ARnine}, are described in section \ref{sec:nxrx}. In section \ref{sec:results} we present the results, including coefficients for production rate evaluation (\ref{sec:empirical}) and an estimate of the uncertainty in the calculated results (\ref{sec:uncert}). A discussion follows in section \ref{sec:discuss} where we compare our calculations with recent results \citep[][sec.\,\ref{sec:compare}]{yatsevich:1997,mei:2010,yokochi:2012,yokochi:2013,yokochi:2014corr}, calculate the nucleogenic neutron flux out of a rock (\ref{sec:nflux}), and discuss the geochemical implications (\ref{sec:geochem}). Concluding remarks follow in section \ref{sec:conclude}.

\section{Overview}
\label{sec:overview}

\begin{figure}
\centering
\includegraphics[width=\linewidth]{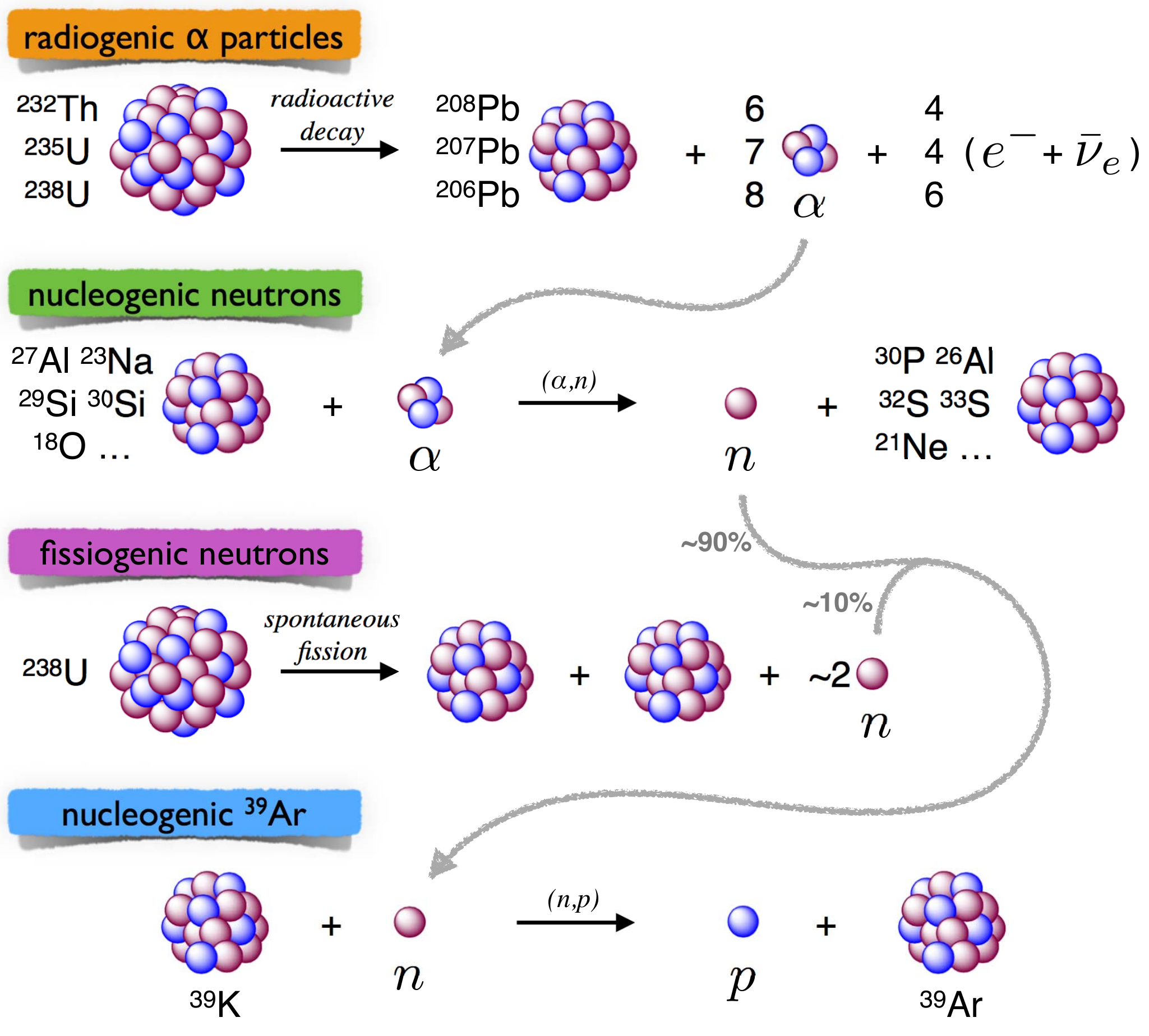}
\caption{Overview of nucleogenic {\ARnine} production.}
\label{fig:schema}
\end{figure}

Decays of natural radionuclides of U and Th in Earth's interior produce $\alpha$ particles with specific known energies. The $\alpha$ particles slow down (i.e., lose kinetic energy) due to collisions with other atoms that form the silicate rock and eventually stop and form {\HEfor} atoms by ionizing nearby matter. A small fraction ($\lesssim10^{-6}$) of the $\alpha$ particles, before losing all their energy, participate in {\an} reactions with light (atomic number $Z\lesssim25$) target nuclides to produce fast ($\sim1$\,MeV) neutrons and product nuclides. Additional neutrons, about 10\,\% of the total neutrons, are produced in spontaneous fission (SF) of {\Ueit} (mean SF neutron energy 1.7\,MeV). Collectively all of these neutrons can then participate in a number of interactions, including scattering and various reactions including {\np} reactions. We are interested in the {\np} reaction on {\Knine} to produce {\ARnine} (Figure~\ref{fig:schema}).

The number of {\ARnine} atoms produced per unit time per unit mass of rock is denoted {\Snine} ($S$ for source). To calculate {\Snine} it is necessary to know the neutron production rate as well as the neutron energy distribution. The neutron spectra are required because reaction cross sections, including that of {\Knine\np\ARnine}, strongly depend on the energy of the incident particles. To calculate {\an} neutron production and spectra, the production rate of $\alpha$ particles and their energy distribution must be known. In sections \ref{sec:alpha} to \ref{sec:nxrx} we discuss the production of $\alpha$ particles, neutrons, and {\ARnine}. A useful byproduct of the {\ARnine} calculation is the production rate of another noble gas nuclide, {\NEone}, which is produced in the {\an} reaction on {\Oeit} with a limited contribution from {\MGfor\na\NEone} reactions.

We evaluate the results for a number of representative rock compositions. Our selection includes the three layers of Continental Crust (CC) of \citet{rudnick:2014tgc}, the Bulk Oceanic Crust (OC) of \citet{white:2014tgc}, and the Depleted Upper Mantle (DM) composition of \citet{salters:2004}. Table~\ref{tab:comp} lists the elemental abundances. We also provide plug-in {formul\ae} allowing inputs of an arbitrary elemental composition to obtain production rates.

\section{Alpha-particle production}
\label{sec:alpha}

\begin{figure}
\centering
\includegraphics[width=\linewidth]{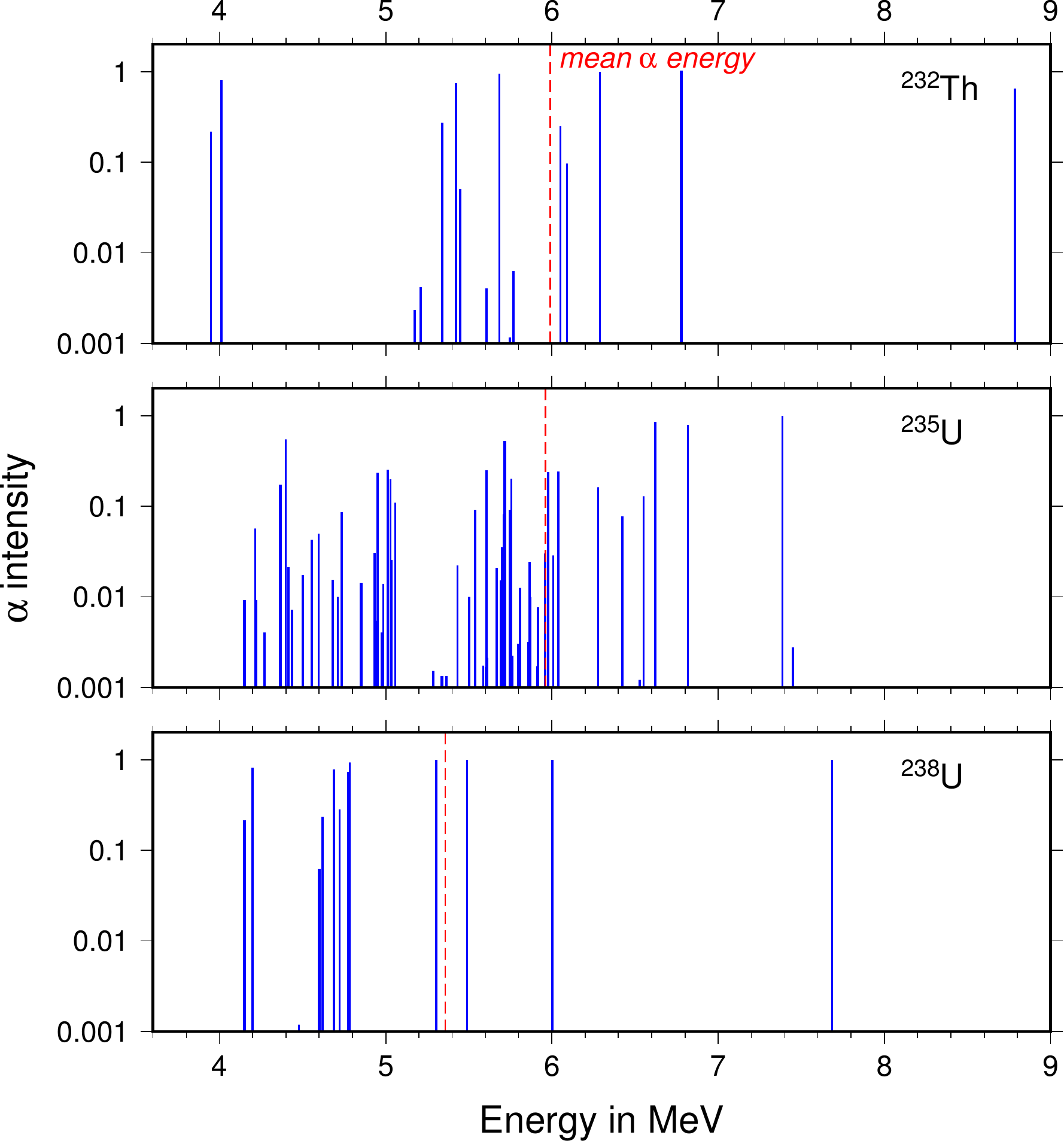}
\caption{Natural $\alpha$ emission energy spectra. Blue spikes represent intensity of each alpha emitted. Spike intensities sum up to number of alphas emitted in particular decay chain (6, 7, 8 for \THtwo, \Ufive, \Ueit, respectively). Red dashed line shows the mean energy of emitted $\alpha$ particles.}
\label{fig:alphaE}
\end{figure}

Alpha particles are produced in radioactive decays of naturally occurring \isot{Th}{232}, \isot{U}{235}, and \isot{U}{238} and their daughter nuclides along the decay chains. Other natural $\alpha$ emitters are not included as their contribution is negligible (the next most potent $\alpha$ emitter, {\isot{Sm}{147}}, gives 20 times fewer $\alpha$ particles per unit mass of rock per unit time than {\Ufive}). Each individual $\alpha$-decay emits one $\alpha$ particle with specific energies, given by the decay scheme which is characterized by the energy levels and intensities. We use decay data---half-lives, branching ratios, $\alpha$ energy levels, $\alpha$ intensities---from ``Chart of Nuclides'' available at the National Nuclear Data Center website \citep{NNDC}. The resulting $\alpha$ energy spectra are shown in Figure~\ref{fig:alphaE}. Tabulated $\alpha$ energies and intensities as well as charts of decay networks can be found in Supplementary Materials.
The maximum $\alpha$ energy is 8.78\,MeV from decay of {\isot{Po}{212}} in the thorium decay chain. The mean $\alpha$ energy in \isot{Th}{232}, \isot{U}{235}, and \isot{U}{238} chains is 6.0, 6.0, and 5.4\,MeV, respectively.

Decays of \isot{Th}{232}, \isot{U}{235}, and \isot{U}{238} produce $n_\alpha=6$, 7, and 8 $\alpha$'s per chain. The $\alpha$ production rate $\Salf$---number of $\alpha$ particles produced per unit time in unit mass of rock---of each decay chain is proportional to the elemental abundance (mass fraction) $A$ of the parent (Th or U) and is obtained by simple multiplication,
\begin{equation} \label{Salf}
\Salf = \nalf A \lambda \frac{XN_A}{M} = \nalf A \cfac,
\end{equation}
where we have introduced the per-decay-to-rate conversion factor $\cfac$,
\begin{equation} \label{cfac}
\cfac \equiv \lambda \frac{XN_A}{M},
\end{equation}
in which $\lambda$ is the decay constant and the fraction $XN_A/M$ is the number of parent nuclide atoms per unit mass of element (e.g.,  number of {\Ueit} atoms per 1\,kg of natural U) where $X$ is natural isotopic composition (mole fraction), $M$ is standard atomic weight, and $N_A$ is Avogadro's number (see Table~\ref{tab:decaypar} for values). The three equations \eqref{Salf}, one for each decay chain, can be recast into one plug-in formula to calculate $\alpha$ production rate from uranium concentration in ppm and Th/U ratio,
\begin{equation} \label{Salfempir}
\Salf \left[\frac{\#}{\text{kg-rock\,s}}\right] = \left( 103 + 24.3 \ThU \right) A_\text{U}\text{[ppm]}.
\end{equation}
Alpha production rates $\Salf$ are evaluated in Table~\ref{tab:alfprod} for the representative rock compositions.

\begin{figure}
\centering
\includegraphics[width=\linewidth]{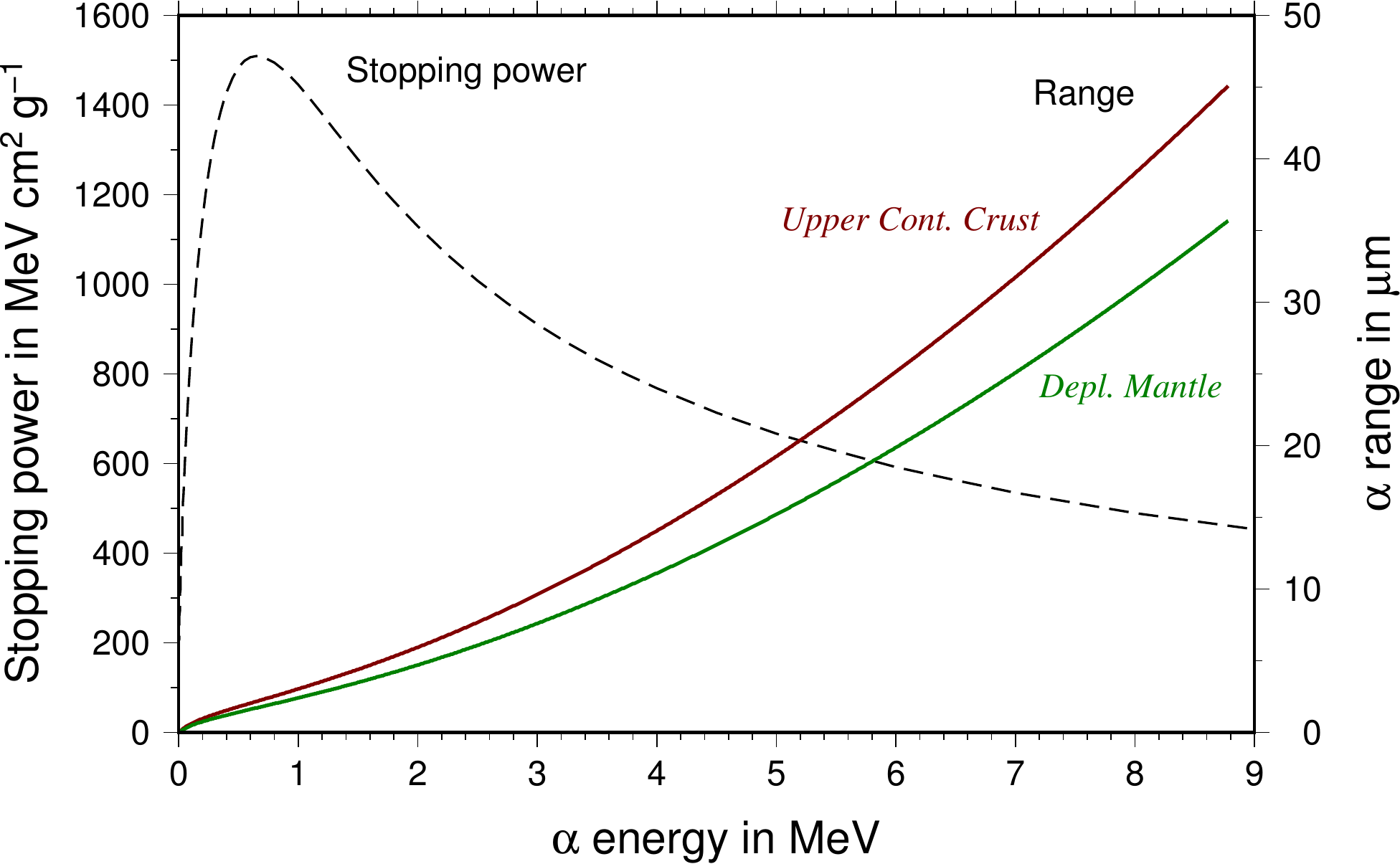}
\caption{Mass stopping power $-\rho^{-1}\tdl{E}{x}$ for an $\alpha$ particle in rock (dashed line, left vertical axis) and range of $\alpha$ particle (eqn.\,\ref{range}, solid lines, right vertical axis) as a function of $\alpha$ energy. Range plotted up to maximum energy of a natural $\alpha$ particle (8.78\,MeV from decay of $\isot{Po}{212}$). The difference between range in Upper Cont. Crust (red) and Depleted Mantle (green) simply reflects the difference in rock density (Table~\ref{tab:comp}). Curves for Middle and Lower CC and OC fall between the two shown.}
\label{fig:sprange}
\end{figure}

Each $\alpha$ particle is emitted with a specific initial kinetic energy {\Eaz} (Figure~\ref{fig:alphaE}). It progressively loses its energy, mostly by inelastic scattering on atomic electrons and elastic scattering on nuclei. The range of an $\alpha$ particle, i.e., how far it travels before stopping, is obtained by integration along its path,
\begin{equation} \label{range}
\text{Range}(\Eaz) = \int\limits_{\Eaz}^0 \drm x = \int\limits_0^{\Eaz} \frac{\drm \Ea}{\left(-\frac{\drm \Ea}{\drm x}\right)},
\end{equation}
and is a function of the initial energy of the $\alpha$ particle in a given material. The material property $-\tdl{\Ea}{x}$ is the linear stopping power. We use the {\srim} software \citep[\href{http://srim.org}{srim.org};][]{ziegler:2008srim}, version {\tt SRIM-2013.00}, to obtain the mass stopping power $-\rho^{-1}\tdl{\Ea}{x}$, where $\rho$ is rock density (listed in Table~\ref{tab:comp}). Mass stopping power is essentially identical for all rock compositions we consider and its energy dependence is plotted in Figure~\ref{fig:sprange}. The range of an average-energy natural $\alpha$ ($\sim6$\,MeV) in the rock is about 25\,$\mu$m (Figure~\ref{fig:sprange}). We assume that $\alpha$ particle propagation is isotropic with no channeling in the crystal lattice on the grain scale.

\section{Neutron production}
\label{sec:neutron}

The overall neutron production rate {\Sneut}---number of neutrons produced per unit time in unit mass of rock---in each decay chain is calculated from
\begin{equation}
\Sneut = A \cfac Y,
\end{equation}
where the neutron yield $Y$, i.e., the number of neutrons produced from decay or fission of one atom of parent nuclide, is the sum of contributions from {\an} reactions on various target nuclides and from spontaneous fission,
\begin{equation}
Y = \sum\limits_i^\text{\#\,targets} \Yan^i + \Ysf.
\end{equation}
We now calculate the individual neutron yields. Energy spectra of the neutrons produced via various production pathways are also discussed in the following sections.

\subsection{Neutrons from {\an} reactions}

Neutron production by {\an} was comprehensively described and quantified by \citet{feige:1968}. Before an $\alpha$ particle loses all its kinetic energy, it can enter another atom's nucleus to form a compound nucleus. The $\alpha$ has to have enough energy to overcome the Coulomb barrier, i.e., the electromagnetic repulsion between the target nucleus and the $\alpha$ particle. The Coulomb barrier height {\Vcoul} is estimated from
\begin{equation} \label{Vcoul}
\Vcoul = \frac{1}{4\pi\epsilon_0} \frac{q_1q_2}{r} = 1.4400 \frac{Z_1 Z_2}{R_{int}\text{[fm]}}\,\text{MeV} = \frac{1.4400\, Z_1 Z_2}{1.2(\mathcal{A}_1^{1/3}+\mathcal{A}_2^{1/3})}\,\text{MeV},
\end{equation}
where indices 1 and 2 refer to projectile and target, $\epsilon_0$ is vacuum permittivity, $q$ is electric change, $r$ is distance, $Z$ is atomic number, $R_{int}$ (in femtometers) is the interaction radius taken as the sum of the two atomic radii, which in turn are approximated by a common empirical formula assuming nuclear volume to be proportional to the mass number $\mathcal{A}$. As the Coulomb barrier height is proportional to the charge of the target nucleus, only relatively low $Z$ target nuclides allow for compound nucleus formation from natural $\alpha$ particles. Even though the compound nucleus can form with $\alpha$ energies below the Coulomb barrier due to quantum tunneling effect, the interaction cross section drops rapidly.

The compound nucleus is considered a short-lived intermediate state and from this compound nucleus there are many possible channels that form various products nuclides. At energies of natural $\alpha$ particles, the most common outcome is that the compound nucleus then sheds a neutron to form a product nucleus. This constitutes an {\an} reaction. For example, $\alpha+\Oeit$ form a compound $\isot{Ne}{22}^*$, which then emits a neutron to produce the {\NEone} product or, in shorthand, {\Oeit\an\NEone}. The majority of the relevant {\an} reactions are endothermic, i.e., their $Q$ value is negative ($Q$ value being reactant minus product rest masses), and the incoming $\alpha$ particle has to have energy above a threshold {\Eth} for the reaction to proceed. Specifically, the kinetic energy in the center-of-mass system before interaction has to be larger than $|Q|$ if $Q<0$. In the laboratory reference frame, this translates into a relation for the threshold energy of an endothermic reaction
\begin{equation} \label{Eth}
\Eth = -\frac{m_1+m_2}{m_2}Q,
\end{equation}
where $m_1$ is the rest mass of projectile and $m_2$ that of target nuclide.

The Coulomb barrier and threshold energy result in a strong energy dependence of the {\an} reaction cross section. Based on the magnitude of cross sections and the abundance of a particular nuclide in natural rocks, we identify 14 target nuclides that are important for {\an} neutron production in natural rocks. They are listed in Table~\ref{tab:targets} together with the {\an} reaction $Q$ values, threshold energies {\Eth} \eqref{Eth}, and Coulomb barrier heights {\Vcoul} \eqref{Vcoul}.

The chance that an $\alpha$ particle, emitted with initial energy $\Eaz$, participates in an {\an} reaction on a light target nuclide $i$ is quantified by the thick target---meaning that the $\alpha$ stops within the medium---neutron production function
\begin{equation} \label{npf}
P_i(\Eaz) = N_i \int\limits_{\Eaz}^0 \frac{\CSan^i(\Ea)}{\td{\Ea}{x}}  \drm \Ea,
\end{equation}
where $N_i$ is the atomic density of nuclide $i$ (\# atoms of $i$ per unit volume) and $\CSan^i$ is the {\an} cross section for nuclide $i$. To get from neutron production function to {\an} neutron yield consists of, for each decay chain and target nuclide $i$, accounting for all $\alpha$ decays and $\alpha$ energy levels within each decay, giving
\begin{equation} \label{Yan}
\Yan^i = \sum\limits_{k=1}^\text{decays} b_k \sum\limits_{l=1}^\text{levels} f_{kl} P_i(E_{kl}),
\end{equation}
where $b_k$ accounts for decay chain branching and $f_{kl}$ is the $\alpha$ intensity of level $l$ in decay $k$ with $\alpha$ energy $E_{kl}$.

To calculate the {\an} neutron energy spectra, we evaluate the differential neutron production function $\tdl{P_i}{E_n}$,
\begin{equation} \label{dnpf}
\td{P_i}{E_n}(\Eaz,E_n) = N_i \int\limits_{\Eaz}^0 \frac{\td{\CSan^i}{E_n}(\Ea,E_n)}{\td{\Ea}{x}} \drm \Ea,
\end{equation}
which is an equation analogous to \eqref{npf} except that it calls for the neutron spectrum (or differential cross section) $\tdl{\CSan^i}{E_n}$, instead of the integrated cross section $\CSan^i$ ($E_n$ denotes neutron energy). To obtain the yield spectrum (or differential yield) $\tdl{\Yan^i}{E_n}$, we follow the analogy through equation \eqref{Yan}. 
Similar thick target methods for $\alpha$ particles have been used to study nuclear reaction rates for astrophysics \citep{roughton:1983}, and these methods have been used for thermonuclear reactions in energetic plasmas \citep{intrator:1981}.

\begin{figure}
\centering
\includegraphics[width=\linewidth]{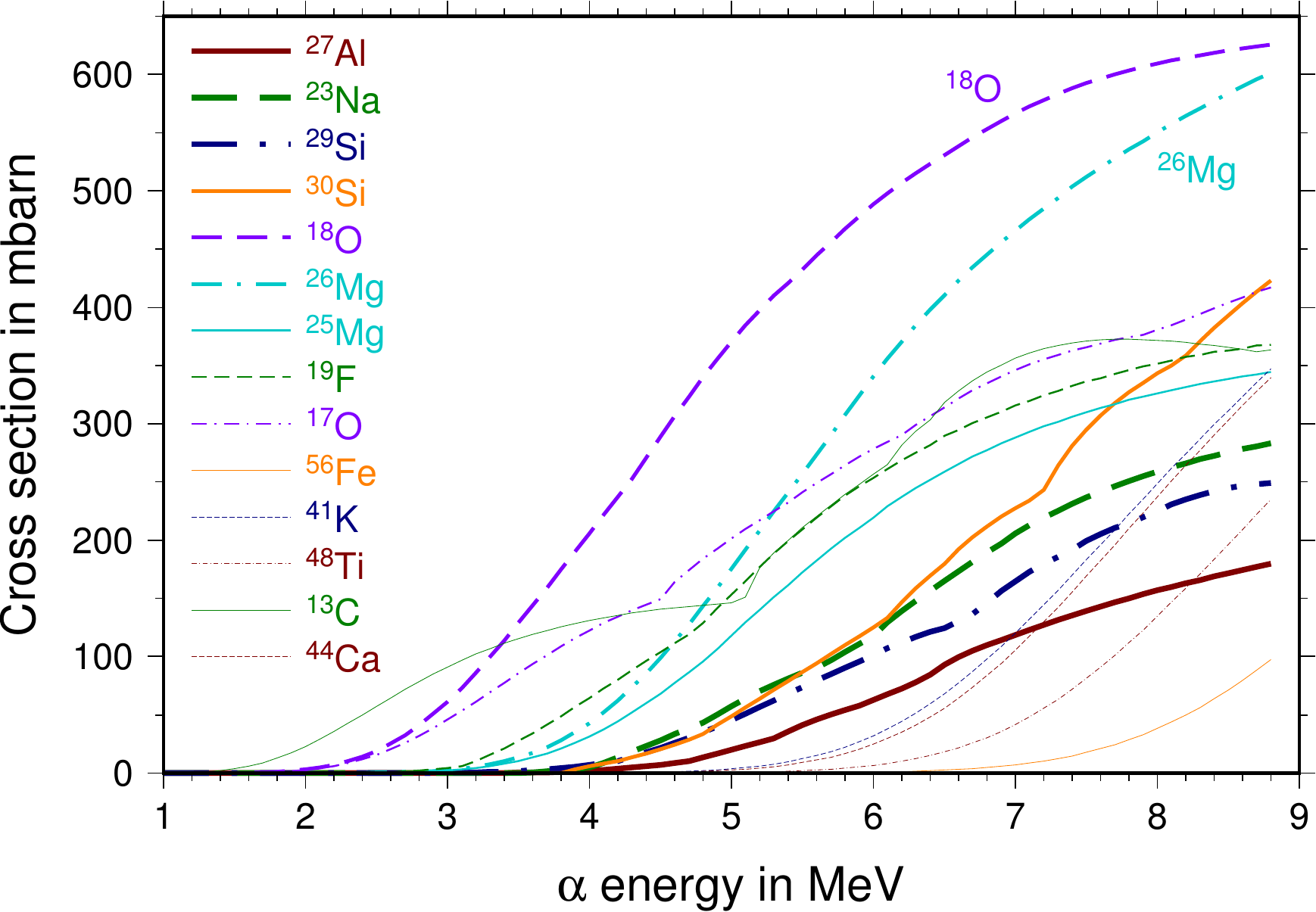}
\caption{Energy dependent {\an} cross sections $\CSan$ for target nuclides considered in this study as calculated by {\talys} version 1.6. One mbarn = $10^{-27}$\,cm$^2$.}
\label{fig:anxs}
\end{figure}

We use the nuclear physics code {\talys} \citep[\href{http://www.talys.eu/}{www.talys.eu};][]{TALYS-1.0}, version 1.6 released on December 23, 2013, to calculate the {\an} neutron production cross sections $\CSan^i$ as well as the emitted neutron energy spectra $\tdl{\CSan^i}{E_n}$. We use the default {\talys} input parameters except for allowing the width fluctuation corrections calculated using the Moldauer model at all energies (see the {\talys} documentation for details). An example of our {\talys} input file is provided in Supplementary Materials.
The {\talys}-calculated {\an} cross sections for all considered target nuclides are plotted in Figure~\ref{fig:anxs}.

\begin{figure}
\centering
\includegraphics[width=\linewidth]{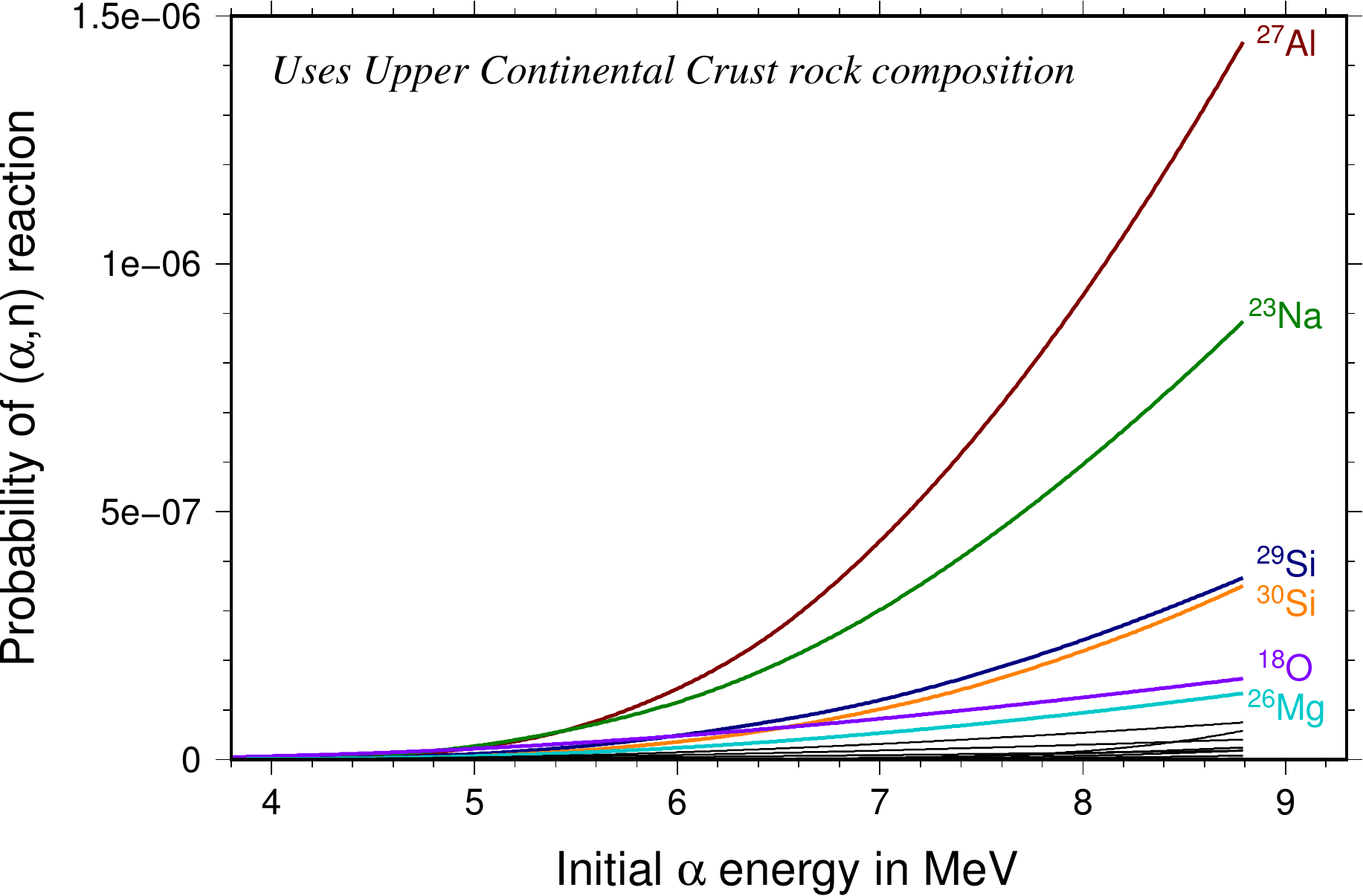}
\caption{Thick target {\an} neutron production function $P$ (i.e., neutron yield per one $\alpha$ particle; eqn.\,\ref{npf}) plotted against the initial $\alpha$ energy for various target nuclides. Curves for the six most neutron producing targets are labeled and color-coded. Remaining targets plotted as thin black lines. The Upper CC rock composition is used. Different amplitudes for various target nuclides stem from the combined effect of elemental abundance, natural isotopic composition, and energy dependence of {\an} cross section.}
\label{fig:npf}
\end{figure}

\begin{figure}
\centering
\includegraphics[width=\linewidth]{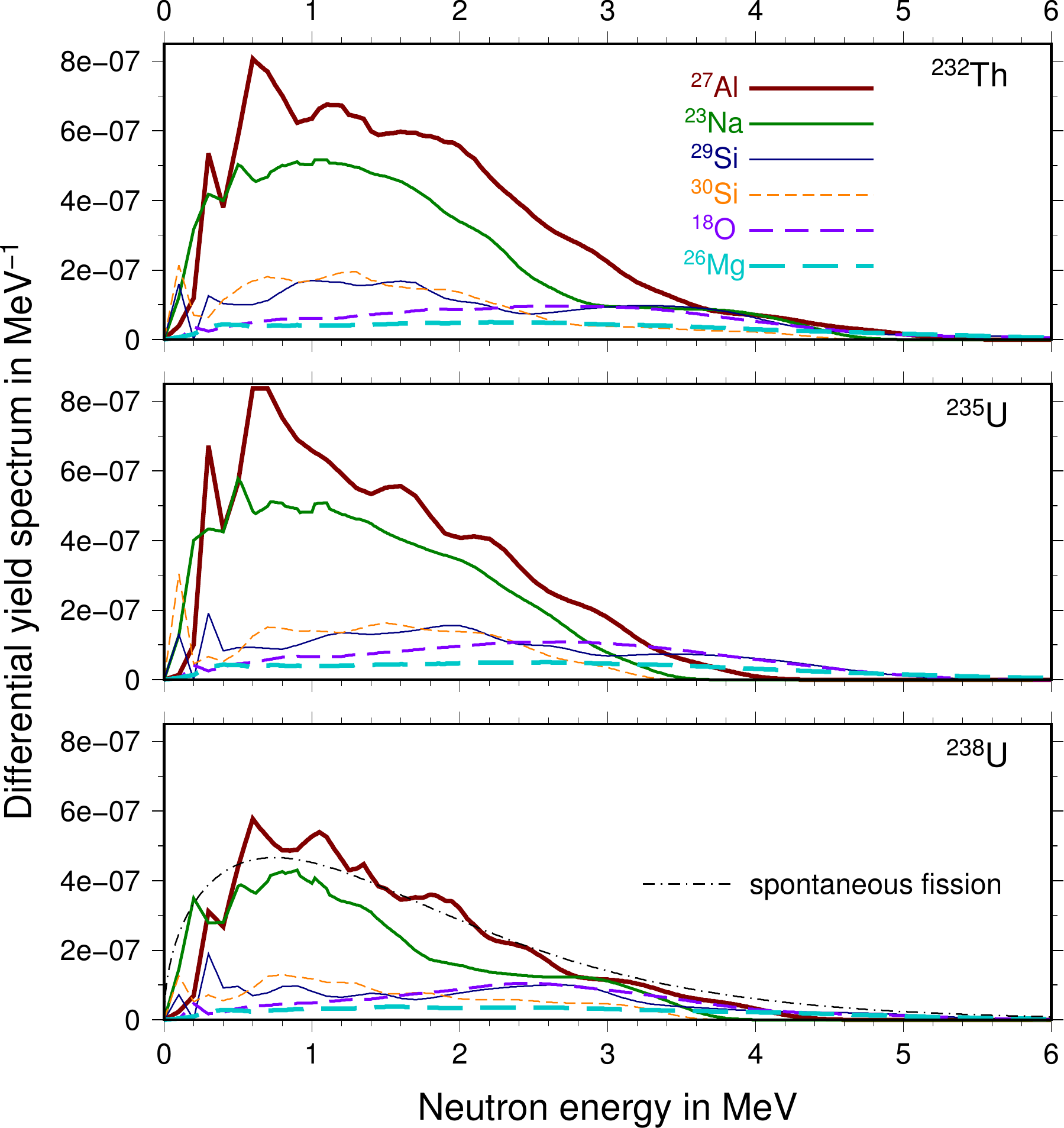}
\caption{{\an} neutron yield spectra $\tdl{\Yan}{E_n}$ plotted against neutron energy for each decay chain and six most neutron producing targets. The fission neutron yield spectrum $\tdl{\Ysf}{E_n}$ is also plotted (non-negligible only in {\Ueit} decay). The Upper CC rock composition is used. Area below each curve integrates to the neutron yield $\Yan$ or $\Ysf$ (i.e., neutrons produced per decay of one atom of parent nuclide).}
\label{fig:dnyield}
\end{figure}

The calculated neutron production function $P_i$ \eqref{npf} of various target nuclides is presented in Figure~\ref{fig:npf}. The figure shows the relative importance of various target nuclides, which depends both on the cross section magnitude and the nuclide abundance in the rock. Figure~\ref{fig:dnyield} shows the neutron yield spectra $\tdl{\Yan^i}{E_n}$ for each decay chain and target nuclide from each of the three actinides. The mean energy of neutrons generated by {\an} reactions is 1.8\,MeV in {\THtwo} decay chain, 1.6\,MeV in {\Ufive} chain, and 1.7\,MeV in {\Ueit} chain.

\subsection{Neutrons from spontaneous fission}

Spontaneous fission of {\Ueit} produces 2.07 neutrons per fission \citep{shultis:2002} and the SF branching ratio is $5.5\times10^{-7}$ \citep{NNDC}. The fission neutron spectrum is approximated by the Watt fission spectrum with neutron energy distribution following $\exp(-E_n/a)\sinh\sqrt{bE_n}$ with parameters $a=0.6483$\,MeV and $b=6.811$\,MeV \citep{SOURCES4C}, resulting in a mean SF neutron energy of 1.7\,MeV. The fission neutron yield spectrum $\tdl{\Ysf}{E_n}$ is included in Figure~\ref{fig:dnyield}. The number of neutrons produced in spontaneous fission of {\THtwo} and {\Ufive}, as well as other nuclides with non-zero SF branching fraction in the decay chains (\isot{U}{234}, \isot{Th}{230}, \isot{Pa}{231}), is negligible compared to neutrons from {\an} reactions.

\section{Neutron-induced reactions: {\np} and {\na}}
\label{sec:nxrx}

To quantify the last step in the nuclear reaction sequence of {\ARnine} production, the {\Knine\np\ARnine} exothermic reaction (Table~\ref{tab:targets}), we use {\mcnp}, version {\tt MCNP6\_Beta3}, a general-purpose Monte Carlo N-Particle transport code developed at Los Alamos National Laboratory (\href{http://mcnp.lanl.gov/}{mcnp.lanl.gov}). {\mcnp} allows us to calculate the {\ARnine} yield per one neutron where we specify the neutron energy spectrum for each of the {\an} and SF neutron production channels. An example of our {\mcnp} input file is provided in Supplementary Materials.
In addition to {\ARnine} yields, we also use {\mcnp} to calculate the {\NEone} yield from the {\isot{Mg}{24}\na\NEone} endothermic reaction (Table~\ref{tab:targets}). From these nuclide yields and neutron production rates discussed in previous section, we can evaluate the {\ARnine} and {\NEone} production rates $\Snine$ and $\Sone$.

\section{Results}
\label{sec:results}

\subsection{Neutron yields and production rates}

The calculated yields $Y$ and production rates $\Sneut$ of neutrons produced by both {\an} reactions and by spontaneous fission are reported in Table~\ref{tab:nyr}. For the Upper CC compositions we report detailed results for each {\an} target nuclide and each natural decay chain to show the relative importance of various neutron production channels. For the remaining representative compositions, only the {\Oeit\an\NEone}, SF, and total neutron yields are presented. The maximum neutron yields are of the order of $3\times10^{-6}$\,neutrons per decay of 1 atom of long-lived radionuclide. With Upper CC composition, spontaneous fission is responsible for 11\,\% of total neutrons produced; with other compositions the SF contribution ranges from 8 to 12\,\%. Of {\an} target nuclides in Upper CC rocks, {\isot{Al}{27}} produces the most neutrons at 32\,\% of total, followed by {\isot{Na}{23}} at 23\,\%, {\isot{Si}{29}} at 9.2\,\%, {\isot{Si}{30}} at 7.8\,\%, {\isot{O}{18}} at 7.0\,\%, {\isot{Mg}{26}} at 4.0\,\% and {\isot{Mg}{25}} at 2.4\,\%. These proportions obviously vary with composition, however, these seven target nuclides + SF neutrons account for at least 96\,\% of neutrons produced for all representative compositions we use. With the magnesium-rich DM composition, Mg nuclides alone account for 69\,\% of neutrons. In terms of contributions of decay chains, with Upper CC composition {\THtwo} accounts for 57\,\% and {\Ueit} for 41\,\% of neutrons produced. Again, exact proportions vary with composition but the trend, $\THtwo>\Ueit\gg\Ufive$, holds.

\subsection{{\ARnine} and {\NEone} yields and production rates}

The largest {\np} and {\na} nuclide yields (per neutron per elemental weight fraction of reacting nuclide, i.e., K or Mg) are obtained with highest energy neutron spectra, coming from exothermic {\an} reactions with positive $Q$ values. That is, from {\an} target nuclides {\isot{C}{13}}, {\isot{O}{17}}, {\isot{Mg}{26}}, and in particular {\isot{Mg}{25}}, which provides the highest energy neutrons. The maximum nuclide yields are of the order of 0.4\,per neutron per weight fraction of K for {\ARnine} and 0.05\,per neutron per weight fraction of Mg for {\NEone}. The production rates, per year per kilogram of rock, also factor in the relative importance of neutron production channels. Therefore, the seven most neutron producing nuclides ({\isot{Al}{27}}, {\isot{Na}{23}}, {\isot{Si}{29}}, {\isot{Si}{30}}, {\isot{O}{18}}, {\isot{Mg}{26}}, {\isot{Mg}{25}}) and SF are also responsible for most of {\ARnine} production. Neon-21 production by {\na} is dominated by {\an} neutrons from {\isot{Mg}{25}} (55--81\,\% of {\NEone} produced depending on chosen composition). The results for {\ARnine} and {\NEone} production rates are reported in Table~\ref{tab:nyr}.

Table~\ref{tab:rates} summarizes the subsurface production of $\alpha$ particles, neutrons, {\NEone}, and {\ARnine}. Depending on selected rock composition, on the order of $10^7$--$10^{10}$ $\alpha$ particles are produced in one kilogram of rock per year. The number of neutrons produced is lower by $\sim6$ orders of magnitude, {\NEone} production rate drops by an additional factor of 15--20, and production rate of {\ARnine} decreases by at least another order of magnitude. In Table~\ref{tab:rates} we include production rates of {\HEfor}, {\NEone}, and {\ARnine} in units of cm$^3$ STP per year per gram of rock, in order to facilitate easier comparison of our results to previous work.

\subsection{Coefficients for plug-in formul\ae}
\label{sec:empirical}

In Table~\ref{tab:coeffs} we provide coefficients $\chi$ that one can use to calculate subsurface neutron production, {\NEone} production by {\an} and nucleonic {\ARnine} production in a rock of arbitrary composition. The word ``arbitrary'' should be qualified with ``while close enough to a natural rock composition''. To indicate what ``close enough'' means, let us state that while the provided coefficients are based on a calculation with Upper CC composition, the empirical formulas yield a result within 1\,\% of the actual value for all rock compositions we use here (Upper, Middle, and Lower CC, Bulk OC, Depleted Upper Mantle). The one exception is {\ARnine} production evaluation for DM composition, where the Upper CC coefficients overestimate {\ARnine} by 8\,\%.

As an example, we use these coefficients to evaluate the neutron and {\ARnine} production rate in the {\Oeit} {\an} target nuclide channel with $\alpha$ particles from {\Ueit} decay chain in a rock with Upper CC composition. The appropriate neutron production coefficient $\chi_n(\Oeit, \Ueit)$ in Table~\ref{tab:coeffs} is $2.27\times10^8$ and the neutron production rate is evaluated, using elemental abundances from Table~\ref{tab:comp}, as
\begin{equation} \label{evaln}
\begin{split}
\Sneut &= \chi_n(\Oeit, \Ueit) ~ \times ~ A_\text{O} ~ \times ~ A_\text{U}\\
 &= 2.27\times10^8 ~ \times ~ 0.480 ~ \times ~ 2.7\times10^{-6}\\
 &= 294\,\text{neutrons/(year\,kg-rock)},
\end{split}
\end{equation}
which, as a consistency check, agrees with the rate reported in Table~\ref{tab:nyr}. 
With {\Oeit} target, the neutron production rate is equal to the {\NEone} production rate by {\an}. 
The {\ARnine} production coefficient $\chi_{39\text{Ar}}(\Oeit, \Ueit)$ in Table~\ref{tab:coeffs} is $4.68\times10^7$ and the {\ARnine} production rate is evaluated as
\begin{equation} \label{evala}
\begin{split}
\Snine &= \chi_{39\text{Ar}}(\Oeit, \Ueit) ~ \times ~ A_\text{O} ~ \times ~ A_\text{U} ~ \times ~ A_\text{K}\\
 &= 4.68\times10^7 ~ \times ~ 0.480 ~ \times ~ 2.7\times10^{-6} ~ \times ~ 0.0232\\
 &= 1.41\,\text{atoms/(year\,kg-rock)},
\end{split}
\end{equation}
which again agrees with Table~\ref{tab:nyr} result.
A Python script which performs the evaluations \eqref{evaln} and \eqref{evala} is provided via \href{http://tinyurl.com/argon39}{tinyurl.com/argon39}.

\subsection{Uncertainty in the calculation}
\label{sec:uncert}

Estimates of rock composition generally carry large uncertainty, especially as one aims to infer the chemistry of the deep Earth \citep[for example, the amount of heat-producing elements K, Th, and U in the Earth is only agreed upon up to a factor of a few; e.g., ][]{mcdonough:2014}. Still, it is worthwhile to estimate the uncertainty of neutron, {\NEone} and {\ARnine} production, assuming source rock composition is known. The additional uncertainty arises from uncertainties in the nuclear physics models we employ and their underlying data sets, i.e., the evaluation of stopping power (\srim), neutron production cross sections (\talys), and neutron interactions (\mcnp).

Half-lives of $\alpha$-decaying nuclides, branching ratios, $\alpha$ energy levels, and $\alpha$ intensities have small uncertainties \citep{NNDC}; some uncertainties are listed in Table~\ref{tab:decaypar}. We estimate the overall uncertainty of calculating $\alpha$ particle production to be below 1\,\%.

\citet{ziegler:2010} report the accuracy of {\srim} stopping power calculations when compared to experimental data, which is 3.5\,\% for stopping of $\alpha$ particles. Since these stopping powers are vital to implementation technologies in the semiconductor industry, these values and uncertainty can be used with confidence.

Estimation of {\an} and {\np} cross section uncertainties are challenging. Neutron reactions on common elements in the few MeV energy range are important to many applications for fission reactors. Therefore one might expect data, models, and codes to have quite small uncertainties. Various nuclear data libraries exist, such as the ENDF/B-VII library (\href{http://www.nndc.bnl.gov/endf/}{www.nndc.bnl.gov/endf}) used by {\mcnp}, which provide nuclear cross section data. The ``evaluated nuclear data'' come from assessment of experimental data combined with nuclear theory modeling. The evaluated library datasets come with no uncertainty estimate, however. Often large differences exist between cross sections from various libraries and without provided uncertainties, any rigorous comparison is impossible. Various experimental datasets are available, some of which include uncertainty estimate. However, some experimental data are at odds with cross sections from nuclear data libraries.

Given this unfortunate situation, we adopt the following strategy for estimating cross section uncertainty. Cross sections used in this study are compared to experimental data available via the EXFOR database at \href{http://www-nds.iaea.org/exfor/}{www-nds.iaea.org/exfor}. We integrate each cross section over energy up to a certain upper energy bound and calculate the relative difference between the integrated data sets. We use this difference as a measure of the cross section uncertainty ($1\sigma$).

Clearly, this method of uncertainty estimation is not satisfying. However, given the lack of rigorous uncertainty estimates provided by existing data libraries, the variability in the cross section data gives us at least some estimate of uncertainty. \citet{koning:2012} describe the method of rigorous uncertainty estimation in {\talys} nuclear data evaluations which consists of a Monte Carlo approach to propagation of uncertainty in various nuclear model parameters. Unfortunately, the available versions of {\tt TENDL} (\talys-based evaluated nuclear data library) do not include the uncertainty in {\an} cross section and angular distribution.

In the case of {\an} reactions, we integrate {\talys}-calculated cross sections, used in this study, and the available experimental dataset (using linear interpolation between the data points) up to 6\,MeV (mean $\alpha$ energy in the {\THtwo} chain). The relative difference between the integrals is taken as $1\sigma$ uncertainty estimate. In the case of the {\Knine\np} reaction, the cross section from ENDF/B-VII.0 library used by {\mcnp} and an experimental dataset are integrated up to 3\,MeV (above which the neutron energy spectrum is negligible). Again, the resulting relative difference is adopted as $1\sigma$ uncertainty.

Table~\ref{tab:error} lists the various contributors to the calculation uncertainty. We considered the seven most important {\an} target nuclides, together with spontaneous fission accounting for $>96\,\%$ of neutron production. Experimental data from \citet{flynn:1978} (EXFOR entires A0509005, A0509007, A0509008) were used for {\isot{Al}{27}}, {\isot{Si}{29}} and {\isot{Si}{30}};  \citet{norman:1982} (C0731001) data for {\isot{Na}{23}}; combined  \citet{bair:1962} (P0120002) and \citet{hansen:1967} (P0116003) data for {\Oeit}. In the case of {\isot{Mg}{25}} and {\isot{Mg}{26}} where no experimental data are available in the EXFOR database, we arbitrarily set the uncertainty at 10\,\%. 
The experimental {\Knine\np} cross section dataset is constructed from \citet{johnson:1967} and \citet{bass:1964} experimental data. \citet{nolte:2006} report a non-zero {\Knine\np\ARnine} cross section in the epithermal energy range. However, the cross section is smaller by at least 4 orders of magnitude compared to values for fast ($\sim1$\,MeV) neutrons, and thus will contribute negligibly to the {\ARnine} production. Plots of cross sections for visual inspection can be found in Supplementary Materials.

The uncertainty of the neutron propagation calculation is represented solely by the {\Knine\np} cross section uncertainty, which is an oversimplification. Strictly speaking, one should consider the uncertainty in cross sections of all reactions that the neutron can participate in. Such complete treatment is beyond this study's scope. However, {\Knine\np} has the largest of all {\np} cross sections \citep{khuukhenkhuu:2011}, which justifies our simple approach. 
The N-particle simulations performed by {\mcnp} also introduce a statistical uncertainty.  Argon-39 yield tallies are repeated until the statistical uncertainty is below 0.5\,\%, therefore negligible compared to systematic uncertainty. Statistical uncertainty of {\NEone} tallies are as high as several tens of percent in some cases. However, as we show, the contribution of {\isot{Mg}{24}\na\NEone} production is negligible relative to {\Oeit\an\NEone} in rocks.

The overall uncertainty of {\ARnine} production is calculated using standard error propagation rules following this symbolic formula for {\ARnine} production rate evaluation,
\begin{equation} 
\label{error}
[\text{{\ARnine}\,production}] = [\alpha\,\text{decay}] \times [\text{stopping\,power}]^{-1} \left\{\sum\limits_\text{target}[\text{\an\,reaction}]\right\} \times [\text{\Knine\np\,reaction}],
\end{equation}
where each term in [brackets] carries its own uncertainty and these uncertainties are considered independent. Evaluation of neutron and {\NEone} production uncertainty is modified accordingly. We evaluate the overall uncertainty of both the neutron production calculation and {\NEone} production calculation at $<20$\,\%; the uncertainty estimate of {\ARnine} production is 30\,\% (Table~\ref{tab:error}). We used the composition of Upper Continental Crust in the uncertainty estimation. The choice of elemental composition is reflected in the relative importance of various neutron production channels which somewhat changes the overall uncertainty with composition. These uncertainty estimates do not include the uncertainty in rock composition.
To our knowledge, this is the first attempt at consistent uncertainty estimation of the {\ARnine} nucleonic production rate. \citet{yokochi:2014corr} state a 50\,\% uncertainty of their {\ARnine} production calculations, however, no reasoning is offered to justify their estimate.

Our calculation assumes a homogeneous elemental distribution in the rock. This is a good assumption for the Earth's mantle. In the crust, however, the presence of accessory mineral phases introduces a heterogeneity in distribution of some elements on the grain size scale, typically a few hundred $\mu$m and particularly so for K, Th \& U. In the Continental Crust the distribution of Th \& U is heavily controlled by accessory phases \citep{bea:1996}. Given the mean free path of a MeV neutron in the rock, a few centimeters, the geometric effect of this heterogeneity on neutron propagation is insignificant. However, as the range of a natural $\alpha$ particle is only few tens of microns, this heterogeneity may change the neutron production both in terms of rate and energy content, depending on how various {\an} target nuclides are spatially distributed within the mineral phases (see Figures~\ref{fig:npf} and \ref{fig:dnyield}). This would in turn affect the neutron induced reaction yields of {\ARnine} and {\NEone}. 
\citet{martel:1990} developed a simple geometric model of spherical accessory phase inclusion in host rock and quantified their effect on neutron yields relative to uniform distribution of elements. In the limit of grain size exceeding the most energetic $\alpha$ particle range (25\,$\mu$m), they find that presence of uraninite and monazite inclusions (deficient in the important light element {\an} targets Al, Na, Si, O) in biotite decreases the neutron production by a factor of 5. The neutron and subsequently {\ARnine} production clearly depends strongly on petrography. 

\section{Discussion}
\label{sec:discuss}

\subsection{Comparison to previous results}
\label{sec:compare}

We provide a comparison of nucleogenic production rates of neutrons, {\NEone}, and {\ARnine} calculated in this study to select recent evaluations \citep{yatsevich:1997,leya:1999,mei:2010,yokochi:2012,yokochi:2014corr}. Similar calculations were previously performed by \citet{fabryka-martin:1988phd,andrews:1989b,andrews:1991,lehmann:1993,lehmann:1997}.

\citet{mei:2010} calculated rates of {\ARnine} production for one representative granitic rock composition. We assume that their reported Th and U abundance values were erroneously interchanged in their manuscript, as we are not aware of any granites where Th to U ratio is $<1$. Using their corrected elemental composition, we calculate a neutron production rate at $5400\pm700$\,neutrons/(kg\,yr), which is identical to the result of 5500\,neutrons/(kg\,yr) calculated by \citet{mei:2009}. The agreement is expected as both \citet{mei:2009} and our calculation use {\an} cross section inputs from {\talys}. \citet{mei:2010} then estimate the {\ARnine} production rate to be 7\,atoms per kg of rock per year. Our calculation gives $16\pm5$\,atoms/(kg\,yr). The factor of two discrepancy stems from different methods of calculating the {\Knine\np\ARnine} reaction yields. \citet{mei:2010} estimate the {\ARnine} production rate as proportional to the ratio of {\Knine\np\ARnine} cross section to total neutron absorption cross section. It is not obvious whether---and how---\citet{mei:2010} account for the energy spectrum on the neutrons generated by {\an} reactions and spontaneous fission. In this work, we use {\mcnp} which calculates large numbers of histories of neutrons distributed along a calculated energy spectrum and accounts for all the possible interactions in the material of given composition. Of course, all these models assume, perhaps incorrectly, a uniform distribution of elements.

\citet{yokochi:2012} calculated nucleogenic {\ARnine} production for a few representative compositions using a ``modified version of SLOWN2 code''. With their Continental Crust composition with 2.0\,\%\,K, 5.0\,ppm\,Th and 1.5\,ppm\,U, their calculated production rate is 0.065\,{\ARnine} atoms/(cm$^3$\,yr) which, assuming rock density 2.7\,g\,cm$^{-3}$, translates to 24\,atoms/(kg\,yr). When we rescale our result for Middle CC composition which is the closest in K, Th, U abundance to their K, Th, U, assuming other elemental abundances unchanged, we arrive at $13\pm4$\,atoms/(kg\,yr), another factor of two difference in results, in the other direction.

\citet{yokochi:2013} calculated the {\ARnine} production rate for two specific rock compositions. Their rates were 4 to 6 times above our calculations, in addition to showing some obvious inconsistencies (a lower {\ARnine} production rate for a composition richer in K, Th, U). Following \citet{sramek:2013agu} and subsequent discussion, \citet{yokochi:2014corr} updated their result for ``Lava Creek Tuff'' rock composition to $120\pm60$\,atoms/(kg\,yr). Our calculation with their K, Th, U input abundances yields $140\pm40$\,atoms/(kg\,yr). Their SLOWN2 code calculation assumes mono-energetic neutrons at 2\,MeV while our calculation accounts for the actual neutron energy spectrum. When we assume all {\an} and SF neutrons are produced with an initial energy of 2\,MeV (while keeping the neutron yields unchanged), we observe a 35\% decrease in the {\ARnine} production rate. \citet{yokochi:2014corr}'s result falls in between our calculation with the full neutron spectra (Figure~\ref{fig:dnyield}) and our calculation with 2\,MeV neutrons, and within the large uncertainty agrees with our result.

Table~\ref{tab:neon} breaks down {\NEone} production into the {\Oeit\an} and {\isot{Mg}{24}\na} channels. The neutron induced {\NEone} production is negligible for crustal compositions and only contributes 3.4\,\% to {\NEone} produced in a Depleted Upper Mantle rock. We calculate the nucleogenic {\NEone/\HEfor} production ratio at $(4.6\pm0.6)\times10^{-8}$ in Continental Crust and at $(4.2\pm0.5)\times10^{-8}$ in Oceanic Crust and Depleted Mantle. Within uncertainty, our result agrees with \citet{yatsevich:1997} who calculated the {\NEone/\HEfor} production ratio at $4.5\times10^{-8}$. Their optimistic uncertainty of $<5\,\%$ comes from the small uncertainty (2\,\%) adopted for the {\an} yield in their analysis. Using chemical abundances for the ``crust'' and ``mantle'' compositions of \citet{mason:1982book}, which were the input concentrations in both \citet{yatsevich:1997} and \citet{leya:1999} studies, we calculate {\NEone} production rates of $469\pm78$\,atoms/(kg\,yr) ($(1.77\pm0.29)\times10^{-20}$\,cm$^3$\,STP/(g\,yr)) for the crust and $4.4\pm0.7$\,atoms/(kg\,yr) ($(1.67\pm0.28)\times10^{-22}$\,cm$^3$\,STP/(g\,yr)) for the mantle. Our results lie in between the lower rates of \citet{leya:1999} and the higher values of \citet{yatsevich:1997}, while both fall within our $1\sigma$ uncertainty bounds; see also \citet{ballentine:2002}.

\subsection{Surface flux of nucleogenic and fissiogenic neutrons}
\label{sec:nflux}

Near the ground, nucleogenic and fissiogenic fast neutrons produced in the rock can be ejected into the atmosphere and enhance the near-surface neutron flux derived from cosmic rays. Most of these neutrons that might be ejected from the rock would originate within the first few tens of centimeters within the surface and from $\sim3$\,meter depth at most. To investigate whether the nucleogenic and fissiogenic fast neutrons may be a significant contribution to the near-surface neutron flux, we have calculated the neutron flux for two specific rock compositions: granite is the median USGS G-2 composition and limestone is the median NIST SRM1c composition (Table~\ref{tab:comp}), both datasets obtained from the GeoReM database (\href{http://georem.mpch-mainz.gwdg.de}{georem.mpch-mainz.gwdg.de}). The calculated neutron flux out of a flat rock surface is $7.8\times10^{-5}$\,{\ncms} for granite and $5.1\times10^{-6}$\,{\ncms} for limestone. The factor of 15 difference reflects the disparate Th and U content of the ``hot'' granite and the ``cold'' limestone. These fluxes only account for the {\an} and SF neutrons. It turns out that their contribution is minor compared to the cosmic ray neutron production, where the flux of 0.4\,eV--0.1\,MeV neutrons was measured by \citet{yamashita:1966} to be $2.9\times10^{-3}$\,{\ncms} at sea level. Cosmic ray neutrons therefore constitute the dominant component ($>99\,\%$) of neutron flux at Earth's surface.

\subsection{Implications for groundwater dating}
\label{sec:geochem}

Given the consequences of global climate change and the world population increase, a need is arising to explore pumping of groundwater from previously unused deep reservoirs. Interestingly, \citet{gleeson:2015} demonstrate that only $\sim6$\,\% (1--17\,\% with uncertainties) of groundwater present at depths down to 2\,km around the globe is ``modern'', i.e., younger than 50\,years. This evaluation is based on tritium (\isot{H}{3}) dating (half-life $12.32\pm0.02$\,years). The next available radiometric tool on the age scale is {\ARnine} (half-life $269\pm3$\,years), which fills the gap between the shorter-lived \isot{H}{3} and also \isot{Kr}{85} ($10.739\pm0.014$\,years), and the longer-lived \isot{C}{14} ($5700\pm30$\,years). 
The number of {\ARnine} dating studies and measurements so far were limited due to the large sample size and analytical requirement of the well established Low-Level Counting (LLC) method \citep{loosli:2005}. However, ongoing technological development efforts, in particular the Atom Trap Trace Analysis \citep[ATTA; e.g.,][]{lu:2014}, now allow for a relatively fast {\ARnine} age analysis of small sample volumes \citep{ritterbusch:2014}, and are expected to increase the number of analyses in the future.
Understanding the relationship between the atmosphere- and subsurface-generated {\ARnine} is thus essential in deep groundwater radiometric dating, in studies of fluid circulation in the Earth's crust, circulation pathways and residence times of ocean water masses, and in dating of deep ice cores.

Hydrological and glaciological dating using {\ARnine} relies on counting {\ARnine} atoms in water/ice samples as {\ARnine} activity predictably decays exponentially, it is assumed, after sequestration of the sample from the atmosphere. Cosmogenic neutron flux sharply decreases with depth in the soil/rock, and therefore, the atmospheric production mechanism $\ARfort(n,2n)\ARnine$ becomes insignificant. However, any other mechanism of ambient {\ARnine} production at depth must be factored in the dating analysis in order to obtain correct ages. In principle, several noble gas nuclide production mechanisms are possible underground. Cosmogenic production includes neutron spallation, thermal neutron capture, negative muon ($\mu^-$) capture, and fast muon induced reactions with nuclides abundant in the rock \citep{niedermann:2002}. \citet{mei:2010} showed that $\mu^-$ capture dominates {\ARnine} production at depths down to 1800\,m.w.e. ($\sim700$\,m depth). At greater depths, the nucleogenic {\Knine\np\ARnine} production dominates.

In most old groundwater, {\ARnine} is below detection limit. This is due to the combination of relatively low production rate of {\ARnine} in rocks, slow release from rock matrix to groundwater, and short half-life. Therefore, {\ARnine} studies for water resource assessment are typically based on simple radioactive decay where the subsurface production does not contribute significantly to the {\ARnine} budget and {\ARnine} is simply a decay tracer \citep[e.g.,][]{delbart:2014,edmunds:2014,mayer:2014,visser:2013,sultenfuss:2011,corcho.alvarado:2007}.
Geochemical sites where {\ARnine} subsurface production becomes important are high temperature environments such as geothermal system \citep{purtschert:2009conf,yokochi:2013,yokochi:2014corr}, highly radioactive environments \citep{andrews:1989b}, or settings with intense water--rock interactions. \citet{yokochi:2012} presented a case study for such high {\ARnine} settings.

To account for the nucleogenic {\ARnine} in groundwaters, \citet{yokochi:2012} proposed a novel method using measurements of argon isotopic composition in the fluid. The method is based on comparison of a measurement to a modeled evolution of {\ARnine/\ARfortS} ratio in an argon-producing rock ({\ARfortS} denotes radiogenic \ARfort). Its age range exceeds the usual limit of several half-lives of {\ARnine} --- after which {\ARnine} concentration assumes a steady-state value in an {\ARnine}-producing rock while {\ARfort} continues to accumulate, thus {\ARnine/\ARfort} keeps evolving. The input in their model are the {\ARnine} and {\ARfort} production rates in the rock, which then loses argon to fluid.
\citet{yokochi:2012} present case studies for two sites with available groundwater argon isotopic composition data, Milk River Sandstone and Stripa Granite. In our effort to reproduce their results, we have identified a mistake in their groundwater age calculation: We are only able to recover \citet{yokochi:2012}'s results if we intentionally divide (instead of multiply) with rock density when recalculating from wt.\% K to number of {\Kfort} atoms per cm$^3$ rock. In Table~\ref{tab:yoko12} we show both \citet{yokochi:2012}'s calculation results and our ``original recalculation'' using their input values and rock density 2.7\,g/cm$^3$ for both Milk River Sandstone and Stripa Granite, chosen to best match their output. We also show the ``corrected recalculation,'' where we appropriately multiply with rock density when recalculating from wt.\% K to number of {\Kfort} atoms per cm$^3$ rock. Fixing the calculation error results in a downward revision of both the lower and upper limits on groundwater age by a factor of $\sim7$ (i.e., rock density squared). The corrected upper age limits for Milk River Sandstone samples no longer exceed the age of the reservoir lithology.

For Milk River Sandstone, \citet{yokochi:2012} use a {\ARnine} production rate of 0.015\,atoms/cm$^3$/yr. Using a rock composition of 2.4\,ppm\,U, 6.3\,ppm\,Th, 1.07\,\%\,K, and 3.91\,\%\,Ca \citep{andrews:1991} we calculate a {\ARnine} production rate of 9.45~atoms/kg/yr or, using a rock density of 2.6\,g/cm$^3$ \citep{andrews:1991}, 0.023~atoms/cm$^3$/yr.
For Stripa Granite, \citet{yokochi:2012} use {\ARnine} production rate of 1.3 atoms/cm$^3$/yr. Using the Stripa Granite composition, including 3.84\,\%K, 33.0\,ppm\,Th, 44.1\,ppm\,U \citep{andrews:1989b}, we calculate a {\ARnine} production rate of 367~atoms/kg/yr or 0.95~atoms/cm$^3$/yr, using a rock density of 2.6\,g/cm$^3$ \citep{andrews:1989b}.
Using these new {\ARnine} production rates, the calculated groundwater ages lie within a factor of 2 of the corrected results of \citet{yokochi:2012}. The lower limits on groundwater age for the two Milk River Sandstone samples, Well 8 and Well 9, are calculated to be 6.1\,kyr and 17\,kyr. The upper limits on age are 22\,Myr and 27\,Myr. Both Stripa Granite samples, NI and VI, have {\ARnine/\ARfortS} above the revised value for the reservoir rock assuming closed system. The upper limits on groundwater age of the Stripa Granite samples are calculated to be 0.056\,Myr and 0.35\,Myr (Table~\ref{tab:yoko12}).

\section{Conclusions}
\label{sec:conclude}

Calculations of subsurface nucleogenic {\ARnine} production are extremely useful in hydrology applications and guide strategies for obtaining argon for particle physics detectors, while {\NEone} production is an integral component of geochemical interpretations of noble gas isotopic composition. We describe and calculate the production of neutrons in {\an} interactions of naturally emitted $\alpha$ particles and the nucleogenic production of noble gas nuclides {\ARnine} and {\NEone}. We use new nuclear physics codes to evaluate {\an} reaction cross sections and spectra (\talys) and to track neutron propagation in the rock (\mcnp). 

Nucleogenic {\ARnine} production rate in a rock previously calculated by \citet{mei:2010} is a factor of two below our result, the calculation of \citet{yokochi:2012} is a factor of two above our calculation, and the \citet{yokochi:2014corr} result lies within our uncertainty margin. We consider the sources of uncertainty in the calculation along with identifying areas that require better constraints. The largest uncertainty comes from uncertainties in the nuclear reaction cross sections, which we estimate from the variability among available experimental data and cross sections from nuclear data libraries. 
The overall calculation uncertainty is estimated to be $\sim30$\,\% for {\ARnine} production, and within 20\,\% for {\NEone} and neutron production.
We expect this uncertainty could be decreased if uncertainty estimates were available for the {\an} cross sections within {\talys} and for the ENDF/B-VII library data used by {\mcnp}.
Our analysis yields a tighter uncertainty estimate compared to the recent study by \citet{yokochi:2014corr} who estimated a large error bar (50\,\%) in the calculated {\ARnine} production rate. 

The {\ARnine} chronometer fills in a gap between \isot{H}{3} and \isot{C}{14}. Conventional radiometric {\ARnine} dating of hydrological reservoirs assumes a simple radioactive decay of initial {\ARnine} content in the groundwater. Non-negligible subsurface {\ARnine} production and/or groundwater--rock interaction in specific environments requires a more refined {\ARnine/\ARfortS} chronometer proposed by \citet{yokochi:2012}. Their original calculation for the fluid samples from the Milk River Sandstone and Stripa Granite reservoirs overestimated the groundwater ages by a uniform factor of 7. An update using our new {\ARnine} productions rates results in groundwater age limits which are a factor of 4--11 times lower compared to the original analysis.

The Supplementary Materials\footnote{Supplementary Materials are provided as a separate PDF document.} document the uncertainty estimate of {\an} and {\np} cross sections, and include details of natural decay chains, examples of our {\talys} and {\mcnp} input files, and a link to {\mycode}, a Fortran\,90 code we wrote, which handles the overall calculation.

\section*{Acknowledgements}

We are grateful to Alice Mignerey and Bill Walters for invaluable discussions on broad topics in nuclear chemistry. Jutta Escher brought our attention to {\talys}. Mike Fensin provided helpful advice on {\tt MCNP} usage via the {\tt mcnp-forum} discussion list. Milan Krti\v{c}ka advised on uncertainty estimate.  C\'ecile Gautheron shared her {\Oeit\an\NEone} cross section. Two reviewers provided thorough assessment and thoughtful comments on the manuscript. We gratefully acknowledge support for this research from NSF EAR 0855791 CSEDI  Collaborative Research: Neutrino Geophysics: Collaboration Between Geology \& Particle Physics.

\section*{Author contributions}

O\v{S} and WFM proposed and conceived the study with SM. O\v{S} conducted all simulations and incorporated models and constraints from WFM and SM. Graduate student LS participated early on in data modeling and interpretation of the results. RJP provided insights into the modeling of uncertainties of the nuclear physics data. The manuscript was written by O\v{S} and all the authors contributed to it. All authors read and approved of the final manuscript.


\begin{thebibliography}{62}
\providecommand{\natexlab}[1]{#1}
\expandafter\ifx\csname urlstyle\endcsname\relax
  \providecommand{\doi}[1]{doi:\discretionary{}{}{}#1}\else
  \providecommand{\doi}{doi:\discretionary{}{}{}\begingroup
  \urlstyle{rm}\Url}\fi

\bibitem[{Acosta-Kane et~al.(2008)}]{acostakane:2008}
Acosta-Kane, D., et~al., 2008.
\newblock Discovery of underground argon with low level of radioactive
  {$^{39}$Ar} and possible applications to {WIMP} dark matter detectors.
\newblock Nucl. Instr. Methods Phys. Res. A 587~(1), 46--51.
\newblock \doi{10.1016/j.nima.2007.12.032}.

\bibitem[{Andrews et~al.(1991)Andrews, Florkowski, Lehmann, and
  Loosli}]{andrews:1991}
Andrews, J.~N., Florkowski, T., Lehmann, B.~E., Loosli, H.~H., 1991.
\newblock Underground production of radionuclides in the {M}ilk {R}iver
  aquifer, {A}lberta, {C}anada.
\newblock Appl. Geochem. 6~(4), 425--434.
\newblock \doi{10.1016/0883-2927(91)90042-N}.
\newblock {S}pecial Issue on ``Dating Very Old Groundwater, {M}ilk {R}iver
  Aquifer''.

\bibitem[{Andrews et~al.(1989)}]{andrews:1989b}
Andrews, J.~N., et~al., 1989.
\newblock The in situ production of radioisotopes in rock matrices with
  particular reference to the {S}tripa granite.
\newblock Geochim. Cosmochim. Acta 53~(8), 1803--1815.
\newblock \doi{10.1016/0016-7037(89)90301-3}.

\bibitem[{Bair and Willard(1962)}]{bair:1962}
Bair, J.~K., Willard, H.~B., 1962.
\newblock Level structure in {Ne$^{22}$} and {Si$^{30}$} from the reactions
  {O$^{18}(\alpha,n)$Ne$^{21}$} and {Mg$^{26}(\alpha,n)$Si$^{29}$}.
\newblock Phys. Rev. 128~(1), 299--304.
\newblock \doi{10.1103/PhysRev.128.299}.

\bibitem[{Ballentine and Burnard(2002)}]{ballentine:2002}
Ballentine, C.~J., Burnard, P.~G., 2002.
\newblock Production, release and transport of noble gases in the continental
  crust.
\newblock Rev. Mineral. Geochem. 47~(1), 481--538.
\newblock \doi{10.2138/rmg.2002.47.12}.

\bibitem[{Bass et~al.(1964)Bass, Fanger, and Saleh}]{bass:1964}
Bass, R., Fanger, U., Saleh, F.~M., 1964.
\newblock Cross sections for the reactions $^{39}${K}$(n,p)^{39}${A} and
  $^{39}${K}$(n,\alpha)^{36}${Cl}.
\newblock Nucl. Phys. 56, 569--576.
\newblock \doi{10.1016/0029-5582(64)90503-6}.

\bibitem[{Bea(1996)}]{bea:1996}
Bea, F., 1996.
\newblock Residence of {REE}, {Y}, {T}h and {U} in granites and crustal
  protoliths; implications for the chemistry of crustal melts.
\newblock J. Petrol. 37~(3), 521--552.
\newblock \doi{10.1093/petrology/37.3.521}.

\bibitem[{Benetti et~al.(2007)}]{benetti:2007}
Benetti, P., et~al., 2007.
\newblock Measurement of the specific activity of {$^{39}$Ar} in natural argon.
\newblock Nucl. Instr. Methods Phys. Res. A 574~(1), 83--88.
\newblock \doi{10.1016/j.nima.2007.01.106}.

\bibitem[{Corcho~Alvarado et~al.(2007)}]{corcho.alvarado:2007}
Corcho~Alvarado, J.~A., et~al., 2007.
\newblock Constraining the age distribution of highly mixed groundwater using
  {$^{39}$Ar}: A multiple environmental tracer ({$^3$H/$^3$He, $^{85}$Kr,
  $^{39}$Ar, and $^{14}$C}) study in the semiconfined fontainebleau sands
  aquifer (france).
\newblock Water Resour. Res. 43~(3), W03427.
\newblock \doi{10.1029/2006WR005096}.

\bibitem[{Delbart et~al.(2014)}]{delbart:2014}
Delbart, C., et~al., 2014.
\newblock Investigation of young water inflow in karst aquifers using
  {SF$_6$--FC--H/He--$^{85}$Kr--$^{39}$Ar} and stable isotope components.
\newblock Appl. Geochem. 50, 164--176.
\newblock \doi{10.1016/j.apgeochem.2014.01.011}.

\bibitem[{Dziewonski and Anderson(1981)}]{dziewonski:1981}
Dziewonski, A.~M., Anderson, D.~L., 1981.
\newblock Preliminary reference {E}arth model.
\newblock Phys. Earth Planet. Int. 25~(4), 297--356.
\newblock \doi{10.1016/0031-9201(81)90046-7}.

\bibitem[{Edmunds et~al.(2014)Edmunds, Darling, Purtschert, and
  Corcho~Alvarado}]{edmunds:2014}
Edmunds, W.~M., Darling, W.~G., Purtschert, R., Corcho~Alvarado, J.~A., 2014.
\newblock Noble gas, {CFC} and other geochemical evidence for the age and
  origin of the bath thermal waters, {UK}.
\newblock Appl. Geochem. 40, 155--163.
\newblock \doi{10.1016/j.apgeochem.2013.10.007}.

\bibitem[{Fabryka-Martin(1988)}]{fabryka-martin:1988phd}
Fabryka-Martin, J.~T., 1988.
\newblock Production of radionuclides in the earth and their hydrogeologic
  significance, with emphasis on chlorine-36 and iodine-129.
\newblock Ph.D. thesis, University of Arizona.

\bibitem[{Feige et~al.(1968)Feige, Oltman, and Kastner}]{feige:1968}
Feige, Y., Oltman, B.~G., Kastner, J., 1968.
\newblock Production rates of neutrons in soils due to natural radioactivity.
\newblock J. Geophys. Res. 73~(10), 3135--3142.
\newblock \doi{10.1029/JB073i010p03135}.

\bibitem[{Flynn et~al.(1978)Flynn, Sekharan, Hiller, Laumer, Weil, and
  Gabbard}]{flynn:1978}
Flynn, D.~S., Sekharan, K.~K., Hiller, B.~A., Laumer, H., Weil, J.~L., Gabbard,
  F., 1978.
\newblock Cross sections and reaction rates for $^{23}${Na}$(p,n)^{23}${Mg},
  $^{27}${Al}$(p,n)^{27}${Si}, $^{27}${Al}$(\alpha,n)^{30}${P},
  $^{29}${Si}$(\alpha,n)^{32}${S}, and $^{30}${Si}$(\alpha,n)^{33}${S}.
\newblock Phys. Rev. C 18~(4), 1566--1576.
\newblock \doi{10.1103/PhysRevC.18.1566}.

\bibitem[{Gleeson et~al.(2015)Gleeson, Befus, Jasechko, Luijendijk, and
  Cardenas}]{gleeson:2015}
Gleeson, T., Befus, K.~M., Jasechko, S., Luijendijk, E., Cardenas, M.~B., 2015.
\newblock The global volume and distribution of modern groundwater.
\newblock Nature Geosci. \doi{10.1038/ngeo2590}.

\bibitem[{Graham(2002)}]{graham:2002}
Graham, D.~W., 2002.
\newblock Noble gas isotope geochemistry of mid-ocean ridge and ocean island
  basalts: Characterization of mantle source reservoirs.
\newblock Rev. Mineral. Geochem. 47~(1), 247--317.
\newblock \doi{10.2138/rmg.2002.47.8}.

\bibitem[{Hansen et~al.(1967)}]{hansen:1967}
Hansen, L.~F., et~al., 1967.
\newblock The $(\alpha,n)$ cross sections on {$^{17}$O} and {$^{18}$O} between
  5 and 12.5 {MeV}.
\newblock Nucl. Phys. A 98~(1), 25--32.
\newblock \doi{10.1016/0375-9474(67)90895-0}.

\bibitem[{Intrator et~al.(1981)Intrator, Peterson, Zaidins, and
  Roughton}]{intrator:1981}
Intrator, T.~P., Peterson, R.~J., Zaidins, C.~S., Roughton, N.~A., 1981.
\newblock Determination of proton spectra by thick target radioactive yields.
\newblock Nucl. Instr. Methods Phys. Res. 188~(2), 347--352.
\newblock \doi{10.1016/0029-554X(81)90514-0}.

\bibitem[{Johnson et~al.(1967)Johnson, Chapman, and Callaghan}]{johnson:1967}
Johnson, P.~B., Chapman, N.~G., Callaghan, J.~E., 1967.
\newblock The absolute cross sections of the {$^{39}$K$(n, p)^{39}$Ar} and
  {$^{39}$K$(n,\alpha)^{36}$Cl} reactions for 2.46 {MeV} neutrons.
\newblock Nucl. Phys. A 94~(3), 617--624.
\newblock \doi{10.1016/0375-9474(67)90436-8}.

\bibitem[{Khuukhenkhuu et~al.(2011)Khuukhenkhuu, Odsuren, Gledenov, and
  Sedysheva}]{khuukhenkhuu:2011}
Khuukhenkhuu, G., Odsuren, M., Gledenov, Y.~M., Sedysheva, M.~V., 2011.
\newblock Statistical model analysis of {$(n,p)$} cross sections averaged over
  the fission neutron spectrum.
\newblock J. Korean Phys. Soc. 59~(2), 851--854.
\newblock \doi{10.3938/jkps.59.851}.

\bibitem[{Koning et~al.(2008)Koning, Hilaire, and Duijvestijn}]{TALYS-1.0}
Koning, A.~J., Hilaire, S., Duijvestijn, M.~C., 2008.
\newblock {TALYS-1.0}.
\newblock In O.~Bersillon, F.~Gunsing, E.~Bauge, R.~Jacqmin, S.~Leray, eds.,
  Proceedings of the International Conference on Nuclear Data for Science and
  Technology, April 22-27, 2007, Nice, France, 211--214. EDP Sciences.

\bibitem[{Koning and Rochman(2012)}]{koning:2012}
Koning, A.~J., Rochman, D., 2012.
\newblock Modern nuclear data evaluation with the {TALYS} code system.
\newblock Nucl. Data Sheets 113~(12), 2841--2934.
\newblock \doi{10.1016/j.nds.2012.11.002}.
\newblock {S}pecial Issue on Nuclear Reaction Data.

\bibitem[{Laske et~al.(2001)Laske, Masters, and Reif}]{crust2.0}
Laske, G., Masters, G., Reif, C., 2001.
\newblock {CRUST\,2.0}, {A} new global crustal model at 2x2 degrees.
\newblock \url{http://igppweb.ucsd.edu/~gabi/crust2.html}.

\bibitem[{Lehmann et~al.(1993)Lehmann, Davis, and
  Fabryka-Martin}]{lehmann:1993}
Lehmann, B.~E., Davis, S.~N., Fabryka-Martin, J.~T., 1993.
\newblock Atmospheric and subsurface sources of stable and radioactive nuclides
  used for groundwater dating.
\newblock Water Resour. Res. 29~(7), 2027--2040.
\newblock \doi{10.1029/93WR00543}.

\bibitem[{Lehmann and Purtschert(1997)}]{lehmann:1997}
Lehmann, B.~E., Purtschert, R., 1997.
\newblock Radioisotope dynamics --- the origin and fate of nuclides in
  groundwater.
\newblock Appl. Geochem. 12~(6), 727 -- 738.
\newblock \doi{10.1016/S0883-2927(97)00039-5}.

\bibitem[{Leya and Wieler(1999)}]{leya:1999}
Leya, I., Wieler, R., 1999.
\newblock Nucleogenic production of {Ne} isotopes in {E}arth's crust and upper
  mantle induced by alpha particles from the decay of {U} and {Th}.
\newblock J. Geophys. Res. 104~(B7), 15439--15450.
\newblock \doi{10.1029/1999JB900134}.

\bibitem[{Loosli(1983)}]{loosli:1983}
Loosli, H.~H., 1983.
\newblock A dating method with {$^{39}$Ar}.
\newblock Earth Planet. Sci. Lett. 63~(1), 51--62.
\newblock \doi{10.1016/0012-821X(83)90021-3}.

\bibitem[{Loosli and Oeschger(1968)}]{loosli:1968}
Loosli, H.~H., Oeschger, H., 1968.
\newblock Detection of {$^{39}$Ar} in atmospheric argon.
\newblock Earth Planet. Sci. Lett. 5, 191--198.
\newblock \doi{10.1016/S0012-821X(68)80039-1}.

\bibitem[{Loosli and Purtschert(2005)}]{loosli:2005}
Loosli, H.~H., Purtschert, R., 2005.
\newblock Rare gases.
\newblock In P.~K. Aggarwal, J.~R. Gat, K.~F.~O. Froehlich, eds., Isotopes in
  the Water Cycle: Past, Present and Future of a Developing Science, chap.~7,
  91--96. Springer Netherlands, Dordrecht.
\newblock \doi{10.1007/1-4020-3023-1_7}.

\bibitem[{Lu et~al.(2014)}]{lu:2014}
Lu, Z.-T., et~al., 2014.
\newblock Tracer applications of noble gas radionuclides in the geosciences.
\newblock Earth Sci. Rev. 138, 196--214.
\newblock \doi{10.1016/j.earscirev.2013.09.002}.

\bibitem[{Martel et~al.(1990)Martel, O'Nions, Hilton, and
  Oxburgh}]{martel:1990}
Martel, D.~J., O'Nions, R.~K., Hilton, D.~R., Oxburgh, E.~R., 1990.
\newblock The role of element distribution in production and release of
  radiogenic helium: the {C}arnmenellis {G}ranite, southwest {E}ngland.
\newblock Chem. Geol. 88~(3--4), 207--221.
\newblock \doi{10.1016/0009-2541(90)90090-T}.

\bibitem[{Mason and Moore(1982)}]{mason:1982book}
Mason, B.~H., Moore, C.~B., 1982.
\newblock Principles of Geochemistry.
\newblock Smith \& Wyllie intermediate geology series. Wiley.

\bibitem[{Mayer et~al.(2014)}]{mayer:2014}
Mayer, A., et~al., 2014.
\newblock A multi-tracer study of groundwater origin and transit-time in the
  aquifers of the {V}enice region ({I}taly).
\newblock Appl. Geochem. 50, 177--198.
\newblock \doi{10.1016/j.apgeochem.2013.10.009}.

\bibitem[{McDonough and {\v S}r{\'a}mek(2014)}]{mcdonough:2014}
McDonough, W.~F., {\v S}r{\'a}mek, O., 2014.
\newblock Neutrino geoscience, news in brief.
\newblock Environ. Earth Sci. 71~(8), 3787--3791.
\newblock \doi{10.1007/s12665-014-3133-9}.

\bibitem[{Mei et~al.(2009)Mei, Zhang, and Hime}]{mei:2009}
Mei, D.-M., Zhang, C., Hime, A., 2009.
\newblock Evaluation of induced neutrons as a background for dark matter
  experiments.
\newblock Nucl. Instr. Methods Phys. Res. A 606~(3), 651--660.
\newblock \doi{10.1016/j.nima.2009.04.032}.

\bibitem[{Mei et~al.(2010)}]{mei:2010}
Mei, D.-M., et~al., 2010.
\newblock Prediction of underground argon content for dark matter experiments.
\newblock Phys. Rev. C 81~(5), 055802.
\newblock \doi{10.1103/PhysRevC.81.055802}.

\bibitem[{Niedermann(2002)}]{niedermann:2002}
Niedermann, S., 2002.
\newblock Cosmic-ray-produced noble gases in terrestrial rocks: Dating tools
  for surface processes.
\newblock Rev. Mineral. Geochem. 47~(1), 731--784.
\newblock \doi{10.2138/rmg.2002.47.16}.

\bibitem[{NIST(2016)}]{NISTatom}
NIST, 2016.
\newblock {Atomic Weights and Isotopic Compositions, NIST Standard Reference
  Database 144, \url{http://www.nist.gov/pml/data/comp.cfm}}.

\bibitem[{NNDC(2016)}]{NNDC}
NNDC, 2016.
\newblock {National Nuclear Data Center, Brookhaven National Laboratory,
  \url{http://www.nndc.bnl.gov/}}.

\bibitem[{Nolte et~al.(2006)}]{nolte:2006}
Nolte, E., et~al., 2006.
\newblock Measurements of fast neutrons in {H}iroshima by use of {$^{39}$Ar}.
\newblock Radiat. Environ. Biophys. 44~(4), 261--271.
\newblock \doi{10.1007/s00411-005-0025-0}.

\bibitem[{Norman et~al.(1982)Norman, Chupp, Lesko, Schwalbach, and
  Grant}]{norman:1982}
Norman, E.~B., Chupp, T.~E., Lesko, K.~T., Schwalbach, P., Grant, P.~J., 1982.
\newblock $^{26\mathrm{g,~m}}${Al} production cross sections from the
  $^{23}${Na}$(\alpha,n)^{26}${Al} reaction.
\newblock Nucl. Phys. A 390~(3), 561--572.
\newblock \doi{10.1016/0375-9474(82)90283-4}.

\bibitem[{Purtschert et~al.(2009)Purtschert, Yokochi, Sturchio, Kharaka, and
  Thordsen}]{purtschert:2009conf}
Purtschert, R., Yokochi, R., Sturchio, N.~C., Kharaka, Y., Thordsen, J., 2009.
\newblock {$^{39}$Ar} measurements on hydrothermal fluids from {Y}ellowstone
  {N}ational {P}ark.
\newblock Geochim. Cosmochim. Acta 73~(13), A983--A1062.
\newblock \doi{10.1016/j.gca.2009.05.012}.

\bibitem[{Ritterbusch et~al.(2014)}]{ritterbusch:2014}
Ritterbusch, F., et~al., 2014.
\newblock Groundwater dating with atom trap trace analysis of {$^{39}$Ar}.
\newblock Geophys. Res. Lett. 41~(19), 6758--6764.
\newblock \doi{10.1002/2014GL061120}.

\bibitem[{Roughton et~al.(1983)Roughton, Intrator, Peterson, Zaidins, and
  Hansen}]{roughton:1983}
Roughton, N.~A., Intrator, T.~P., Peterson, R.~J., Zaidins, C.~S., Hansen,
  C.~J., 1983.
\newblock Thick-target measurements and astrophysical thermonuclear reaction
  rates: Alpha-induced reactions.
\newblock Atom. Data Nucl. Data Tables 28~(2), 341--353.
\newblock \doi{10.1016/0092-640X(83)90021-9}.

\bibitem[{Rudnick and Gao(2014)}]{rudnick:2014tgc}
Rudnick, R.~L., Gao, S., 2014.
\newblock Composition of the continental crust.
\newblock In R.~L. Rudnick, ed., The Crust, vol.~4 of \emph{Treatise on
  Geochemistry}, chap.~1, 1--51. Elsevier, Oxford, second edn.
\newblock \doi{10.1016/B978-0-08-095975-7.00301-6}.
\newblock Editors-in-chief H. D. Holland and K. K. Turekian.

\bibitem[{Salters and Stracke(2004)}]{salters:2004}
Salters, V. J.~M., Stracke, A., 2004.
\newblock Composition of the depleted mantle.
\newblock Geochem. Geophys. Geosyst. 5~(5), Q05B07.
\newblock \doi{10.1029/2003GC000597}.

\bibitem[{Shultis and Faw(2002)}]{shultis:2002}
Shultis, J.~K., Faw, R.~E., 2002.
\newblock Fundamentals of Nuclear Science and Engineering.
\newblock CRC Press.
\newblock \doi{10.1201/9780203910351}.

\bibitem[{{SOURCES-4C}(2002)}]{SOURCES4C}
{SOURCES-4C}, 2002.
\newblock {RSICC} {C}omputer {C}ode {C}ollection, {CCC-661}, {C}ode System for
  Calculating $(\alpha,n)$, Spontaneous Fission, and Delayed Neutron Sources
  and Spectra.
\newblock Oak Ridge National Laboratory.

\bibitem[{{\v S}r{\'a}mek et~al.(2013){\v S}r{\'a}mek, McDonough, Mukhopadhyay,
  Stevens, and Siegel}]{sramek:2013agu}
{\v S}r{\'a}mek, O., McDonough, W.~F., Mukhopadhyay, S., Stevens, L., Siegel,
  J., 2013.
\newblock Calculating subsurface nucleonic production of noble gas nuclides:
  Implications on crustal and mantle {K}, {Th}, {U} abundances.
\newblock abstract MR43A-2371 presented at 2013 Fall Meeting, AGU, San
  Francisco, Calif., 9-13 Dec.

\bibitem[{S{\"u}ltenfu{\ss} et~al.(2011)S{\"u}ltenfu{\ss}, Purtschert, and
  F{\"u}hrb{\"o}ter}]{sultenfuss:2011}
S{\"u}ltenfu{\ss}, J., Purtschert, R., F{\"u}hrb{\"o}ter, J.~F., 2011.
\newblock Age structure and recharge conditions of a coastal aquifer (northern
  {G}ermany) investigated with {$^{39}$Ar, $^{14}$C, $^3$H, He} isotopes and
  {Ne}.
\newblock Hydrogeol. J. 19~(1), 221--236.
\newblock \doi{10.1007/s10040-010-0663-4}.

\bibitem[{Tucker and Mukhopadhyay(2014)}]{tucker:2014}
Tucker, J.~M., Mukhopadhyay, S., 2014.
\newblock Evidence for multiple magma ocean outgassing and atmospheric loss
  episodes from mantle noble gases.
\newblock Earth Planet. Sci. Lett. 393, 254--265.
\newblock \doi{10.1016/j.epsl.2014.02.050}.

\bibitem[{Visser et~al.(2013)Visser, Broers, Purtschert, S{\"u}ltenfu{\ss}, and
  de~Jonge}]{visser:2013}
Visser, A., Broers, H.~P., Purtschert, R., S{\"u}ltenfu{\ss}, J., de~Jonge, M.,
  2013.
\newblock Groundwater age distributions at a public drinking water supply well
  field derived from multiple age tracers ({$^{85}$Kr, $^3$H/$^3$He, and
  $^{39}$Ar}).
\newblock Water Resour. Res. 49~(11), 7778--7796.
\newblock \doi{10.1002/2013WR014012}.

\bibitem[{White and Klein(2014)}]{white:2014tgc}
White, W.~M., Klein, E.~M., 2014.
\newblock Composition of the oceanic crust.
\newblock In R.~L. Rudnick, ed., The Crust, vol.~4 of \emph{Treatise on
  Geochemistry}, chap.~13, 457--496. Elsevier, Oxford, second edn.
\newblock \doi{10.1016/B978-0-08-095975-7.00315-6}.
\newblock Editors-in-chief H. D. Holland and K. K. Turekian.

\bibitem[{Xu et~al.(2012)}]{xucalaprice:2012arxiv}
Xu, J., et~al., 2012.
\newblock A study of the residual {$^{39}$Ar} content in argon from underground
  sources.
\newblock ArXiv:1204.6011.

\bibitem[{Yamashita et~al.(1966)Yamashita, Stephens, and
  Patterson}]{yamashita:1966}
Yamashita, M., Stephens, L.~D., Patterson, H.~W., 1966.
\newblock Cosmic-ray-produced neutrons at ground level: Neutron production rate
  and flux distribution.
\newblock J. Geophys. Res. 71~(16), 3817--3834.
\newblock \doi{10.1029/JZ071i016p03817}.

\bibitem[{Yatsevich and Honda(1997)}]{yatsevich:1997}
Yatsevich, I., Honda, M., 1997.
\newblock Production of nucleogenic neon in the {E}arth from natural
  radioactive decay.
\newblock J. Geophys. Res. 102~(B5), 10291--10298.
\newblock \doi{10.1029/97JB00395}.

\bibitem[{Yokochi et~al.(2012)Yokochi, Sturchio, and Purtschert}]{yokochi:2012}
Yokochi, R., Sturchio, N.~C., Purtschert, R., 2012.
\newblock Determination of crustal fluid residence times using nucleogenic
  $^{39}${A}r.
\newblock Geochim. Cosmochim. Acta 88, 19--26.
\newblock \doi{10.1016/j.gca.2012.04.034}.

\bibitem[{Yokochi et~al.(2013)}]{yokochi:2013}
Yokochi, R., et~al., 2013.
\newblock Noble gas radionuclides in {Y}ellowstone geothermal gas emissions: A
  reconnaissance.
\newblock Chem. Geol. 339, 43--51.
\newblock \doi{10.1016/j.chemgeo.2012.09.037}.

\bibitem[{Yokochi et~al.(2014)}]{yokochi:2014corr}
Yokochi, R., et~al., 2014.
\newblock Corrigendum to ``{N}oble gas radionuclides in {Y}ellowstone
  geothermal gas emissions: a reconnaissance'' [{C}hem. {G}eol. 339 (2013)
  43-51].
\newblock Chem. Geol. 371, 128--129.
\newblock \doi{10.1016/j.chemgeo.2014.02.004}.

\bibitem[{Ziegler et~al.(2008)Ziegler, Biersack, and
  Ziegler}]{ziegler:2008srim}
Ziegler, J.~F., Biersack, J.~P., Ziegler, M.~D., 2008.
\newblock SRIM -- {T}he Stopping and Range of Ions in Matter.
\newblock SRIM Company.

\bibitem[{Ziegler et~al.(2010)Ziegler, Ziegler, and Biersack}]{ziegler:2010}
Ziegler, J.~F., Ziegler, M.~D., Biersack, J.~P., 2010.
\newblock {SRIM} -- {T}he stopping and range of ions in matter (2010).
\newblock Nucl. Instr. Methods Phys. Res. B 268~(11--12), 1818--1823.
\newblock \doi{10.1016/j.nimb.2010.02.091}.

\end{thebibliography}

\clearpage

\begin{table}
\caption{Composition, as elemental weight fractions ($A$ in g/g), of geochemical units used in this study. Elements with atomic fractions $>10^{-4}$ are included (otherwise entered as --) in addition to all of Th, U, and {\an} target elements, irrespective of their abundances. Continental Crust (CC) composition from \citet{rudnick:2014tgc}, Bulk Oceanic Crust (OC) from \citet{white:2014tgc}, Depleted Mantle (DM) from \citet{salters:2004}. Granite is the median USGS\,G-2 composition from GeoReM database. Limestone is the median NIST\,SRM1c composition from GeoReM. Non-zero amount of carbon was included in the CC and OC composition (0.5\,\%, 0.1\,\%, and 0.02\,\% of CO$_2$ by weight for Upper, Middle, and Lower CC, and 0.05\,wt\% CO$_2$ for OC). Rock density is taken from CRUST2.0 \citep{crust2.0} for the crustal layers and from PREM \citep{dziewonski:1981} for the depleted mantle (average from MOHO to 400\,km depth). Elemental abundances not provided in a particular compositional estimate entered as n/a.}
\label{tab:comp}
\centering
\begin{tabular}{cclllllll}
\hline
$Z$ & symbol & Upper CC & Middle CC & Lower CC & Bulk OC & DM & Granite & Limestone \\
\hline
3 & Li & -- & -- & -- & -- & -- & 3.40E-05 & n/a \\
6 & C & 1.36E-03 & 2.73E-04 & 5.46E-05 & 1.36E-04 & 1.37E-05 & 2.30E-04 & 0.110 \\
7 & N & 8.30E-05 & n/a & -- & n/a & -- & -- & n/a \\
8 & O & 0.480 & 0.469 & 0.452 & 0.446 & 0.440 & 0.494 & 0.488 \\
9 & F & 5.57E-04 & 5.24E-04 & 5.70E-04 & n/a & 1.10E-05 & 1.28E-03 & n/a \\
11 & Na & 0.0243 & 0.0251 & 0.0197 & 0.0164 & 2.15E-03 & 0.0300 & 1.88E-04 \\
12 & Mg & 0.0150 & 0.0216 & 0.0437 & 0.0621 & 0.230 & 4.72E-03 & 2.50E-03 \\
13 & Al & 0.0815 & 0.0794 & 0.0894 & 0.0831 & 0.0227 & 0.0815 & 3.40E-03 \\
14 & Si & 0.311 & 0.297 & 0.250 & 0.234 & 0.210 & 0.318 & 0.0320 \\
15 & P & 6.55E-04 & 6.55E-04 & 4.36E-04 & 4.36E-04 & -- & 6.11E-04 & 1.75E-04 \\
16 & S & -- & -- & 3.45E-04 & n/a & -- & -- & n/a \\
17 & Cl & 3.70E-04 & 1.82E-04 & 2.50E-04 & n/a & -- & -- & n/a \\
19 & K & 0.0232 & 0.0191 & 5.06E-03 & 6.51E-04 & 6.00E-05 & 0.0370 & 2.40E-03 \\
20 & Ca & 0.0257 & 0.0375 & 0.0685 & 0.0843 & 0.0250 & 0.0130 & 0.359 \\
22 & Ti & 3.84E-03 & 4.14E-03 & 4.91E-03 & 6.59E-03 & 7.98E-04 & 2.84E-03 & 4.20E-04 \\
24 & Cr & -- & -- & -- & 3.17E-04 & 2.50E-03 & -- & -- \\
25 & Mn & 7.74E-04 & 7.74E-04 & 7.74E-04 & 8.52E-04 & 1.05E-03 & -- & -- \\
26 & Fe & 0.0385 & 0.0460 & 0.0655 & 0.0635 & 0.0627 & 0.0185 & 2.88E-03 \\
28 & Ni & -- & -- & -- & -- & 1.96E-03 & -- & -- \\
38 & Sr & -- & -- & -- & -- & -- & 4.75E-04 & -- \\
56 & Ba & -- & -- & -- & -- & -- & 1.88E-03 & -- \\
90 & Th & 1.05E-05 & 6.5E-06 & 1.2E-06 & 2.10E-07 & 1.37E-08 & 2.47E-05 & 9.49E-07 \\
92 & U & 2.7E-06 & 1.3E-06 & 2E-07 & 7.00E-08 & 4.70E-09 & 1.95E-06 & 1.42E-06 \\
\hline
\multicolumn{2}{l}{K/U} & 8609 & 14687 & 25320 & 9300 & 12766 & 18974 & 1690 \\
\multicolumn{2}{l}{Th/U} & 3.9 & 5.0 & 6.0 & 3.0 & 2.9 & 13 & 0.67 \\
\hline
\multicolumn{2}{l}{$\rho$ in g\,cm$^{-3}$} & 2.70 & 2.88 & 3.05 & 2.83 & 3.42 & 2.7 & 2.5 \\
\hline
\end{tabular}
\end{table}%

\begin{table}[h!]
\caption{Atomic and decay quantities for natural long-lived parent nuclides. Natural isotopic composition $X$ and relative atomic mass $M$ from \citet{NISTatom}, half-lives $t_{1/2}$ from \citet{NNDC}, $\lambda=\ln{2}/t_{1/2}$. Atomic mass truncated to six digits. Numbers in parentheses give uncertainty in last digits, otherwise uncertainty beyond shown digits. The conversion factor {\cfac} (eqn.\,\ref{cfac}) is the number of decays per unit time per unit mass of parent element.}
\label{tab:decaypar}
\centering
\begin{tabular}{llllll}
\hline
Quantity & Symbol & Unit & \THtwo & \Ufive & \Ueit \\
\hline
Elemental abundance & $A$ & kg-elem/kg-rock &  &  &  \\
Natural isotopic composition & $X$ & mol-nucl/mol-elem & 1.0000 & 0.007204(6) & 0.992742(10) \\
Standard atomic weight & $M$ & g\,mol$^{-1}$ & 232.038 & 238.029 & 238.029 \\
Half-life & $t_{1/2}$ & $10^9$\,yr & 14.0(1) & 0.704(1) & 4.468(3) \\
Decay constant & $\lambda$ & 10$^{-18}$\,s$^{-1}$ & 1.57(1) & 31.20(4) & 4.916(3) \\
Number of $\alpha$'s per decay chain & $n_\alpha$ &  & 6 & 7 & 8 \\
Per-decay-to-rate conversion & $\cfac$ & $10^6$\,kg-elem$^{-1}$\,s$^{-1}$ & 4.072(29) & 0.5759(9) & 12.35(1) \\
\hline
\end{tabular}
\end{table}

\begin{table}
\caption{Calculated production rates {\Salf} of $\alpha$ particles, as number of $\alpha$'s produced per second in 1\,kilogram of rock, in a particular decay chain (columns 2--4) and the total (column 5).}
\label{tab:alfprod}
\centering
\begin{tabular}{l|lll|l}
\hline
Composition & \THtwo & \Ufive & \Ueit & Total \\
\hline
Upper Continental Crust & 256 & 10.9 & 267 & 533 \\
Middle Continental Crust & 158 & 5.24 & 128 & 292 \\
Lower Continental Crust & 29.2 & 0.806 & 19.8 & 49.8 \\
Bulk Oceanic Crust & 5.11 & 0.282 & 6.91 & 12.3 \\
Depleted Mantle & 0.334 & 0.0190 & 0.464 & 0.817 \\
\hline
\end{tabular}
\end{table}

\begin{table}
\caption{Target and product nuclides, $Q$ values, threshold energies {\Eth} (eqn.\,\ref{Eth}), and Coulomb barriers {\Vcoul} (eqn.\,\ref{Vcoul}) of {\an} reactions. Also $Q$ and {\Eth} of neutron induced reactions. In cases where the product nuclide is unstable$^*$, we show the final stable nuclide as well. Energies in MeV. Nuclide rest masses for calculation of $Q$ and {\Eth} taken from \citet{NISTatom}.}
\label{tab:targets}
\centering
\begin{tabular}{cccccc}
\hline
Target & Reaction & Product & $Q$ & $\Eth$ & $\Vcoul$ \\
\hline
{\isot{Al}{27}} & {\an} & $\isot{P}{30}^* \rightarrow \isot{Si}{30}$ & $-2.6425$ & 3.0345 & 6.8012 \\
{\isot{Na}{23}} & {\an} & $\isot{Al}{26}^* \rightarrow \isot{Mg}{26}$ & $-2.9659$ & 3.4823 & 5.9577 \\
{\isot{Si}{29}} & {\an} & {\isot{S}{32}} & $-1.5258$ & 1.7365 & 7.2107 \\
{\isot{Si}{30}} & {\an} & {\isot{S}{33}} & $-3.4933$ & 3.9598 & 7.1571 \\
{\isot{O}{18}} & {\an} & {\isot{Ne}{21}} & $-0.6961$ & 0.851 & 4.5626 \\
{\isot{Mg}{26}} & {\an} & {\isot{Si}{29}} & $0.0341$ & -- & 6.3298 \\
{\isot{Mg}{25}} & {\an} & {\isot{Si}{28}} & $2.6536$ & -- & 6.3838 \\
{\isot{F}{19}} & {\an} & $\isot{Na}{22}^* \rightarrow \isot{Ne}{22}$ & $-1.9513$ & 2.3624 & 5.0754 \\
{\isot{O}{17}} & {\an} & {\isot{Ne}{20}} & $0.5867$ & -- & 4.6168 \\
{\isot{Fe}{56}} & {\an} & $\isot{Ni}{59}^* \rightarrow \isot{Co}{59}$ & $-5.0961$ & 5.4607 & 11.5272 \\
{\isot{K}{41}} & {\an} & $\isot{Sc}{44}^* \rightarrow \isot{Ca}{44}$ & $-3.3894$ & 3.7206 & 9.0555 \\
{\isot{Ti}{48}} & {\an} & $\isot{Cr}{51}^* \rightarrow \isot{V}{51}$ & $-2.6853$ & 2.9094 & 10.1118 \\
{\isot{C}{13}} & {\an} & {\isot{O}{16}} & $2.2156$ & -- & 3.656 \\
{\isot{Ca}{44}} & {\an} & {\isot{Ti}{47}}  & $-2.1825$ & 2.3812 & 9.3791 \\
\hline
{\isot{K}{39}} & {\np} & {\isot{Ar}{39}}  & $0.2176$ & -- & n/a \\
\hline
{\isot{Mg}{24}} & {\na} & {\isot{Ne}{21}}  & $-2.5554$ & 2.6628 & n/a \\
\hline
\end{tabular}
\end{table}

\begin{landscape}
\begin{table}
\caption{Calculated results for neutron yield $Y$, neutron production rate $\Sneut$, {\ARnine} production rate {\Snine}, and {\NEone} production rate by {\na} reaction {$\Sone^{n,\alpha}$}. {\NEone} production rate by {\an} reaction is equal to neutron production rate with {\Oeit} target. Detailed results---yields or rates from each neutron production channel in each decay chain---are reported for Upper CC rock composition. Only selected output is listed for other compositions. Neutron yields are given per decay of 1 atom of long-lived radionuclide. Production rates given per year per kilogram of rock.}
\label{tab:nyr}
\centering
\small
\begin{tabular}{l|lll|rrrr|rrrr|rrrr}
\hline \hline
 & \multicolumn{3}{c|}{Neutron yield ($Y$)} & \multicolumn{4}{c|}{Neutron production rate (\Sneut)} & \multicolumn{4}{c|}{{\ARnine} production rate (\Snine)} & \multicolumn{4}{c}{{\NEone} prod. rate by {\na} ($\Sone^{n,\alpha}$)} \\
\hline
target & \THtwo & \Ufive & \Ueit & \THtwo & \Ufive & \Ueit & Sum & \THtwo & \Ufive & \Ueit & Sum & \THtwo & \Ufive & \Ueit & Sum   \\
\hline \hline
\multicolumn{4}{l|}{\bf Upper Continental Crust} &  &  &  &  &  &  &  &  &  &  &  &  \\
\isot{Al}{27} & 1.69e-6 & 1.48e-6 & 1.05e-6 & 2265.0 & 72.8 & 1107.0 & 3445.0 & 5.15 & 0.12 & 2.00 & 7.27 & 0.0016 & 0.0000 & 0.0000 & 0.0016 \\
\isot{Na}{23} & 1.15e-6 & 1.07e-6 & 7.66e-7 & 1547.0 & 52.5 & 805.6 & 2405.0 & 2.99 & 0.07 & 1.22 & 4.27 & 0.0003 & 0.0000 & 0.0000 & 0.0003 \\
\isot{Si}{29} & 4.74e-7 & 4.32e-7 & 3.13e-7 & 636.9 & 21.2 & 328.7 & 986.9 & 2.42 & 0.08 & 1.26 & 3.77 & 0.0089 & 0.0000 & 0.0017 & 0.0107 \\
\isot{Si}{30} & 4.09e-7 & 3.50e-7 & 2.53e-7 & 549.2 & 17.2 & 266.0 & 832.4 & 1.04 & 0.03 & 0.49 & 1.55 & 0.0001 & 0.0000 & 0.0000 & 0.0001 \\
\Oeit & 3.28e-7 & 3.51e-7 & 2.80e-7 & 441.4 & 17.2 & 294.2 & 752.8 & 2.30 & 0.09 & 1.41 & 3.80 & 0.0082 & 0.0001 & 0.0018 & 0.0100 \\
\isot{Mg}{26} & 2.01e-7 & 1.99e-7 & 1.43e-7 & 270.0 & 9.8 & 150.1 & 429.8 & 1.43 & 0.05 & 0.76 & 2.24 & 0.0131 & 0.0003 & 0.0051 & 0.0185 \\
\isot{Mg}{25} & 1.18e-7 & 1.19e-7 & 8.53e-8 & 158.1 & 5.8 & 89.8 & 253.7 & 1.00 & 0.04 & 0.59 & 1.64 & 0.0544 & 0.0021 & 0.0304 & 0.0869 \\
\isot{F}{19} & 6.95e-8 & 7.18e-8 & 5.36e-8 & 93.4 & 3.5 & 56.4 & 153.3 & 0.24 & 0.01 & 0.11 & 0.36 & 0.0001 & 0.0000 & 0.0000 & 0.0001 \\
\isot{O}{17} & 3.56e-8 & 3.77e-8 & 3.04e-8 & 47.9 & 1.8 & 31.9 & 81.6 & 0.27 & 0.01 & 0.17 & 0.46 & 0.0028 & 0.0001 & 0.0011 & 0.0039 \\
\isot{Fe}{56} & 3.86e-8 & 7.11e-9 & 9.45e-9 & 51.9 & 0.3 & 9.9 & 62.1 & 0.06 & 0.00 & 0.00 & 0.07 & 0.0000 & 0.0000 & 0.0000 & 0.0000 \\
\isot{K}{41} & 1.99e-8 & 1.14e-8 & 9.81e-9 & 26.7 & 0.6 & 10.3 & 37.6 & 0.06 & 0.00 & 0.02 & 0.08 & 0.0000 & 0.0000 & 0.0000 & 0.0000 \\
\isot{Ti}{48} & 1.30e-8 & 4.65e-9 & 4.92e-9 & 17.5 & 0.2 & 5.2 & 22.9 & 0.06 & 0.00 & 0.01 & 0.07 & 0.0001 & 0.0000 & 0.0000 & 0.0001 \\
\isot{C}{13} & 3.85e-9 & 4.12e-9 & 3.49e-9 & 5.2 & 0.2 & 3.7 & 9.0 & 0.04 & 0.00 & 0.03 & 0.08 & 0.0034 & 0.0001 & 0.0023 & 0.0059 \\
\isot{Ca}{44} & 5.97e-9 & 3.14e-9 & 2.82e-9 & 8.0 & 0.2 & 3.0 & 11.2 & 0.02 & 0.00 & 0.01 & 0.03 & 0.0001 & 0.0000 & 0.0000 & 0.0001 \\
SF & 2.35e-11 & 1.30e-10 & 1.14e-6 & 0.0 & 0.0 & 1198.0 & 1198.0 & 0.00 & 0.00 & 2.97 & 2.97 & 0.0000 & 0.0000 & 0.0211 & 0.0211 \\
\hline
Total &  &  &  & 6119 & 203 & 4360 & 10680 & 17.1 & 0.49 & 11.1 & 28.7 & 0.093 & 0.0027 & 0.064 & 0.159 \\
\hline \hline
\multicolumn{4}{l|}{\bf Middle Continental Crust} &  &  &  &  &  &  &  &  &  &  &  &  \\
\Oeit &  &  &  & 268.2 & 8.1 & 139.0 & 415.4 &  &  &  &  &  &  &  &  \\
SF &  &  &  & 0.0 & 0.0 & 576.6 & 576.7 &  &  &  &  &  &  &  &  \\
\hline
Total &  &  &  & 3876 & 100 & 2137 & 6114 & 9.1 & 0.21 & 4.6 & 13.9 & 0.108 & 0.0025 & 0.054 & 0.165 \\
\hline \hline
\multicolumn{4}{l|}{\bf Lower Continental Crust} &  &  &  &  &  &  &  &  &  &  &  &  \\
\Oeit &  &  &  & 48.1 & 1.2 & 20.8 & 70.1 &  &  &  &  &  &  &  &  \\
SF &  &  &  & 0.0 & 0.0 & 88.7 & 88.7 &  &  &  &  &  &  &  &  \\
\hline
Total &  &  &  & 766 & 17 & 347 & 1129 & 0.52 & 0.010 & 0.21 & 0.75 & 0.074 & 0.0015 & 0.029 & 0.104 \\
\hline \hline
\multicolumn{4}{l|}{\bf Bulk Oceanic Crust} &  &  &  &  &  &  &  &  &  &  &  &  \\
\Oeit &  &  &  & 8.2 & 0.4 & 7.1 & 15.8 &  &  &  &  &  &  &  &  \\
SF &  &  &  & 0.0 & 0.0 & 31.1 & 31.1 &  &  &  &  &  &  &  &  \\
\hline
Total &  &  &  & 133 & 5.8 & 121 & 260 & 0.0126 & 0.00051 & 0.0104 & 0.0235 & 0.0253 & 0.0011 & 0.0189 & 0.0452 \\
\hline \hline
\multicolumn{4}{l|}{\bf Depleted Upper Mantle} &  &  &  &  &  &  &  &  &  &  &  &  \\
\Oeit &  &  &  & 0.53 & 0.03 & 0.47 & 1.03 &  &  &  &  &  &  &  &  \\
SF &  &  &  & 0.00 & 0.00 & 2.09 & 2.09 &  &  &  &  &  &  &  &  \\
\hline
Total &  &  &  & 11.4 & 0.54 & 10.5 & 22.4 & 0.00014 & 0.000007 & 0.00011 & 0.00026 & 0.0207 & 0.0010 & 0.0150 & 0.0366 \\
\hline \hline
\end{tabular}
\end{table}
\end{landscape}

\begin{table}
\caption{Summary of production rates of {\HEfor}, neutrons, {\NEone}, and {\ARnine} in units of both the number of atoms or neutrons per year per kilogram of rock and cm$^3$ STP per year per gram of rock. Compositional estimates are described in Table~\ref{tab:comp}.}
\label{tab:rates}
\centering
\begin{tabular}{l|crrr|ccc}
\hline
 & \multicolumn{4}{c|}{\# of atoms (neutrons) yr$^{-1}$ kg$^{-1}$} & \multicolumn{3}{c}{cm$^3$ STP yr$^{-1}$ g$^{-1}$} \\
Composition & \HEfor & neutrons & \NEone & \ARnine & \HEfor & \NEone & \ARnine \\
\hline
Upper Continental Crust & $1.64\times10^{10}$ & 10680 & 753 & 28.7 & $6.20\times10^{-13}$ & $2.84\times10^{-20}$ & $1.08\times10^{-21}$ \\
Middle Continental Crust & $8.98\times10^{9}$ & 6114 & 416 & 13.9 & $3.39\times10^{-13}$ & $1.57\times10^{-20}$ & $5.25\times10^{-22}$ \\
Lower Continental Crust & $1.53\times10^{9}$ & 1129 & 70.2 & 0.749 & $5.79\times10^{-14}$ & $2.65\times10^{-21}$ & $2.82\times10^{-23}$ \\
Bulk Continental Crust & $9.43\times10^9$ & 6253 & 433 & 15.3 & $3.56\times10^{-13}$ & $1.63\times10^{-20}$ & $5.75\times10^{-22}$ \\
Bulk Oceanic Crust & $3.79\times10^{8}$ & 260 & 15.8 & 0.0235 & $1.43\times10^{-14}$ & $5.96\times10^{-22}$ & $8.87\times10^{-25}$ \\
Depleted Upper Mantle & $2.51\times10^{7}$ & 22.4 & 1.06 & 0.000257 & $9.49\times10^{-16}$ & $4.01\times10^{-23}$ & $9.68\times10^{-27}$ \\
\hline
\end{tabular}
\end{table}

\begin{table}
\caption{Coefficients for evaluation of neutron production (columns 3--5) and {\ARnine} production (columns 6--8) for arbitrary composition. Second column and third row indicate the elements whose abundances ($A$ as weight fractions) cross-multiply a particular coefficient. Neutron production coefficients in units of ``neutrons per year per kg-rock per wt-frac-target-elem per wt-frac-chain-parent-elem''. Argon-39 production coefficients in units of ``{\ARnine} atoms per year per kg-rock per wt-frac-target-elem per wt-frac-chain-parent-elem per K-wt-frac''. See section \ref{sec:empirical} for examples.}
\label{tab:coeffs}
\centering
\begin{tabular}{c||c|ccc|ccc}
\hline
 & & \multicolumn{3}{c|}{Neutron production} & \multicolumn{3}{c}{{\ARnine} production} \\
\hline
 & Chain & \THtwo & \Ufive & \Ueit & \THtwo & \Ufive & \Ueit \\
\hline \hline
{\an} target & $A$ & Th & U & U & Th & U & U \\
\hline
\isot{Al}{27} & Al & 2.65e+9 & 3.31e+8 & 5.03e+9 & 2.59e+8 & 2.36e+7 & 3.92e+8 \\
\isot{Na}{23} & Na & 6.07e+9 & 8.02e+8 & 1.23e+10 & 5.04e+8 & 4.46e+7 & 8.02e+8 \\
\isot{Si}{29} & Si & 1.95e+8 & 2.53e+7 & 3.91e+8 & 3.19e+7 & 4.05e+6 & 6.46e+7 \\
\isot{Si}{30} & Si & 1.68e+8 & 2.05e+7 & 3.17e+8 & 1.37e+7 & 1.35e+6 & 2.51e+7 \\
\Oeit & O & 8.77e+7 & 1.33e+7 & 2.27e+8 & 1.97e+7 & 2.89e+6 & 4.68e+7 \\
\isot{Mg}{26} & Mg & 1.72e+9 & 2.41e+8 & 3.72e+9 & 3.92e+8 & 5.43e+7 & 8.13e+8 \\
\isot{Mg}{25} & Mg & 1.01e+9 & 1.44e+8 & 2.22e+9 & 2.75e+8 & 4.04e+7 & 6.33e+8 \\
\isot{F}{19} & F & 1.60e+10 & 2.34e+9 & 3.75e+10 & 1.78e+9 & 2.06e+8 & 3.17e+9 \\
\isot{O}{17} & O & 9.50e+6 & 1.43e+6 & 2.47e+7 & 2.34e+6 & 3.39e+5 & 5.70e+6 \\
\isot{Fe}{56} & Fe & 1.28e+8 & 3.35e+6 & 9.55e+7 & 6.81e+6 & 2.53e+4 & 1.62e+6 \\
\isot{K}{41} & K & 1.10e+8 & 8.92e+6 & 1.64e+8 & 1.06e+7 & 4.72e+5 & 1.22e+7 \\
\isot{Ti}{48} & Ti & 4.35e+8 & 2.20e+7 & 5.00e+8 & 5.88e+7 & 2.29e+6 & 5.89e+7 \\
\isot{C}{13} & C & 3.61e+8 & 5.48e+7 & 9.96e+8 & 1.31e+8 & 2.10e+7 & 4.02e+8 \\
\isot{Ca}{44} & Ca & 2.98e+7 & 2.22e+6 & 4.29e+7 & 3.96e+6 & 2.13e+5 & 4.80e+6 \\
SF & 1 & 3.01e+3 & 2.37e+3 & 4.44e+8 & 2.88e+2 & 3.09e+2 & 4.73e+7 \\
\hline \hline
\end{tabular}
\end{table}

\begin{table}
\caption{Estimate of calculation uncertainty and its various contributions, assuming chemical composition is known precisely. Contribution of the specific {\an} channel and spontaneous fission to neutron and {\ARnine} production is shown in columns 3 and 4. Calculated with Upper CC composition. $^\dagger$\,No experimental data available, uncertainty was arbitrarily set at 10\,\%.}
\label{tab:error}
\centering
\begin{tabular}{llll}
\hline
 & Uncert. est. & Neutron & \ARnine \\
 & \% & \% contrib. & \% contrib. \\
\hline
Decay data, $\alpha$ production & $<1$ \\
Stopping power & 3.5 \\
~~~~{\isot{Al}{27}\an} cross section & ~~~~36 & 32 & 25 \\
~~~~{\isot{Na}{23}\an} cross section & ~~~~7.7 & 23 & 15 \\
~~~~{\isot{Si}{29}\an} cross section & ~~~~7.3 & 9.2 & 13 \\
~~~~{\isot{Si}{30}\an} cross section & ~~~~20 & 7.8 & 5.4 \\
~~~~{\Oeit\an} cross section & ~~~~17 & 7.0 & 13 \\
~~~~{\isot{Mg}{26}\an} cross section & ~~~~$10^\dagger$ & 4.0 & 7.8 \\
~~~~{\isot{Mg}{25}\an} cross section & ~~~~$10^\dagger$ & 2.4 & 5.7 \\
~~~~Spontaneous fission & ~~~~1 & 11 & 10 \\
Overall {\an}, neutron production & 12 \\
Overall {\an}, {\ARnine} production & 10 \\
{\Knine\np} cross section & 28 \\
\hline
Neutron production calculation & 13 \\
{\NEone} production calculation & 17 \\
{\ARnine} production calculation & 30 \\
\hline
\end{tabular}
\end{table}

\begin{table}
\caption{Calculated production rates of {\NEone} by {\an} and {\na} and {\NEone/\HEfor} ratio. Compositional estimates are described in Table~\ref{tab:comp}.}
\label{tab:neon}
\centering
\begin{tabular}{l|lll|ll|l}
\hline
Composition & \multicolumn{3}{c|}{{\NEone} prod. in atoms/kg-yr} & \multicolumn{2}{c|}{\% contrib.} &  \\
& \an & \na & Total & \an & \na & \NEone/\HEfor \\
\hline
Upper Continental Crust & 753 & 0.159 & 753 & 99.98 & 0.02 & $4.59\times10^{-8}$ \\
Middle Continental Crust & 415 & 0.165 & 416 & 99.96 & 0.04 & $4.63\times10^{-8}$ \\
Lower Continental Crust & 70.1 & 0.104 & 70.2 & 99.85 & 0.15 & $4.58\times10^{-8}$ \\
Bulk Oceanic Crust & 15.8 & 0.0452 & 15.8 & 99.71 & 0.29 & $4.17\times10^{-8}$ \\
Depleted Upper Mantle & 1.03 & 0.0366 & 1.06 & 96.55 & 3.45 & $4.23\times10^{-8}$ \\
\hline
\end{tabular}
\end{table}

\begin{table}[h]
\caption{Update of \citet{yokochi:2012}'s Table 2, using their notation. {\em Original recalculation:} We are only able to recover \citet{yokochi:2012}'s results if we intentionally divide (instead of multiply) with rock density when recalculating from wt.\% K to number of {\Kfort} atoms per cm$^3$ rock. We use \citet{yokochi:2012}'s input values and rock density 2.7\,g/cm$^3$ for both Milk River Sandstone and Stripa Granite. {\em Corrected recalculation:} Appropriately, we multiply with rock density when recalculating from wt.\% K to number of {\Kfort} atoms per cm$^3$ rock. {\em Recalculation with new input:} We use our new values of {\ARnine} production rate {\Pnine} and rock density of 2.6\,g/cm$^3$ for both Milk River Sandstone \citep{andrews:1991} and Stripa Granite \citep{andrews:1989b}. The ratio {\RR} (defined as $\RR \equiv \ARnine/\ARfortS$ in the rock; {\RF} is the ratio in the fluid) is expressed in units of atmospheric $\Ra = 8.1\times10^{-16}$; {\A} is the argon loss constant; {\phiR} is porosity of the rock; Age$_\text{low}$ and Age$_\text{up}$ are the lower and upper limits on the groudwater age.}
\centering
\small
\begin{tabular}{lccc|cccc}
\hline
 & \multicolumn{3}{c|}{INPUT} & \multicolumn{4}{c}{CALCULATION} \\
\hline
 & Rock age & K & \Pnine & {\RR} closed & Age$_\text{low}$ & {\A/\phiR} & Age$_\text{up}$ \\
 & Myr & wt.\% & at./cm$^3$/yr & \Ra & kyr & year$^{-1}$ & Myr \\
\hline\hline
\bf Milk River Sandstone Well 8 & & & & \\
\citet{yokochi:2012} values & 83.5 & 1.1 & 0.015 & 193 & 27.1 & 7.3e-4 & 118 \\
Original recalculation & " & " & " & 192 & 25.5 & 7.2e-4 & 107 \\
Corrected recalculation & " & " & " & 26.4 & 3.5 & 7.2e-4 & 14.7 \\
Recalculation with new input & " & " & 0.023 & 45.7 & 6.1 & 4.3e-4 & 21.9 \\
\hline
\bf Milk River Sandstone Well 9 & & & & \\
\citet{yokochi:2012} values & " & 1.1 & 0.015 & 193 & 75.2 & 3.2e-4 & 328 \\
Original recalculation & " & " & " & 192 & 71.3 & 3.1e-4 & 299 \\
Corrected recalculation & " & " & " & 26.4 & 9.8 & 3.1e-4 & 41.0 \\
Recalculation with new input & " & " & 0.023 & 45.7 & 16.9 & 1.9e-4 & 26.7 \\
\hline\hline
\bf Stripa Granite NI & & & & \\
\citet{yokochi:2012} values & 1650 & 3.8 & 1.3 & 155 & $\RF>\RR$ & 1.3e-3 & 0.45\\
Original recalculation & " & " & " & 153 & $\RF>\RR$ & 1.3e-3 & 0.39 \\
Corrected recalculation & " & " & " & 20.9 & $\RF>\RR$ & 1.3e-3 & 0.054 \\
Recalculation with new input & " & " & 0.95 & 15.9 & $\RF>\RR$ & 1.7e-3 & 0.056 \\
\hline
\bf Stripa Granite VI & & & & \\
\citet{yokochi:2012} values & " & 3.8 & 1.3 & 155 & 3.3 & 2.9e-3 & 2.5 \\
Original recalculation & " & " & " & 153 & 3.1 & 2.9e-4 & 2.5 \\
Corrected recalculation & " & " & " & 20.9 & 0.43 & 2.9e-4 & 0.34 \\
Recalculation with new input & " & " & 0.95 & 15.9 & $\RF>\RR$ & 4.0e-4 & 0.35 \\
\hline
\end{tabular}
\label{tab:yoko12}
\end{table}

\clearpage

\end{document}